\documentclass{article}
\usepackage{amsmath,amsfonts,amsthm}
\usepackage{graphicx}
\begin{document}

\title{\bf Neutrino physics at accelerators}
\author{Enrique Fernandez\thanks{Talk given at the Corfu Summer Institute on Elementary Particle Physics 2005}\\
UAB/IFAE, \\ Campus UAB, Edif Cn, \\08193 Bellaterra, Barcelona, Spain }
\maketitle

\begin{abstract}
Present and future neutrino experiments at accelerators are mainly concerned with understanding the neutrino oscillation phenomenon and its implications. Here a brief account of neutrino oscillations is given together with a description of the supporting data. Some current and planned accelerator neutrino experiments are also explained.
\end{abstract}

\section{Introduction}

Neutrinos are particles that interact only weakly and this gives them a somewhat special character. The cross section for neutrino interactions with matter, for the neutrino energies attainable with accelerators and for normal matter densities, is extremely small, making them enormously penetrating. As a consequence neutrino experiments are difficult, requiring very intense beams and/or very large targets. Since recently we know of another peculiar property of the neutrinos: their mass is extremely small compared to that of the other elementary particles.

Experiments with accelerator neutrinos have played an important role in the past. The discovery of the second and third neutrinos, the discovery of neutral currents in the seventies, and the study of deep inelastic neutrino nucleon scattering, which shed light on the nucleon structure complementary to the information obtained from deep inelastic charged lepton (electron or muon) collisions with nucleons, are some of the highlights of accelerator neutrino physics. The clear evidence for neutrino oscillations first shown by the Super-Kamiokande collaboration in 1998, with the consequence that the neutrinos have a finite mass, has now produced a renaissance of accelerator neutrinos experiments, with a focus on understanding the oscillation phenomenon.

In this lecture a brief account is given of the physics of oscillations and of the experiments that support it, before turning to the current and foreseen oscillation experiments at accelerators. For the studies of neutrino physics at non-accelerators refer to the lectures of D. Wark in this conference.

\section{Neutrino oscillations in vacuum}

In the SM there are 3 lepton families, each containing a charged lepton and a neutrino.
Neutrinos are massless particles and each family lepton number (as well as global lepton number) is conserved.
The neutrino has three states (weak eigenstates): $\nu_e$, $\nu_{\mu}$ and $\nu_{\tau}$. By definition these are the states that couple to the $W$ together with the corresponding charged lepton in charged-current weak interactions. These assumptions, in particular the massless assumption, were built up from experiment.

If neutrinos have mass, the mass eigenstates do not need to be the same as the weak eigenstates. The mass eigenstates are usually denoted by $\nu_1$, $\nu_2$ and $\nu_3$. The situation is similar to that of the quarks, where the weak eigenstates are the Cabbibo-rotated mass states that constitute the hadrons. The difference is that in the neutrino case we do not have a direct access to the mass eigenstates, since we produce and detect neutrinos through the weak interaction, and therefore we project them into pure weak eigenstates when we observe them.

If mass and weak states are not the same they are related by a transformation
\begin{equation}
| \nu_l > = \sum_{i=1}^n U_{li} | \nu_i >
\end{equation}
\noindent where $l$ stands for charged lepton type $e$, $\mu$ or $\tau$, and $i=1,2,3$ refers to mass state. If there are only three mass states, then $U$ is a $3 \times 3$ unitary matrix. Here we consider only this case for the sake of simplicity in the explanation of the oscillations. All the experimental data can be explained with three mass states, except for the results of the LSND experiment \cite{LNSD} (see below). For an excellent review of neutrino oscillations see article of B. Kayser \cite{kayser}. 
The $3 \times 3$ $U$ matrix can be written as follows
\begin{equation}
U = \left ( \begin{array}{ccc} U_{e1} & U_{e2} & U_{e3} \\
U_{\mu 1} & U_{\mu 2} & U_{\mu 3} \\
U_{\tau 1} & U_{\tau 2} & U_{\tau 3} \end{array} \right)
\end{equation}
\noindent This mixing matrix is usually called the Maki-Nakagawa-Sakata-Pontecorvo (MNSP) matrix \cite{MNSP}, analogous to the CKM matrix for the quarks.

To understand what are neutrino oscillations let us assume that a neutrino is produced at $t=0$ in a pure weak eigenstate, $|\nu_{\alpha}>$. This state is a mixture of mass eigenstates determined by the mixing matrix. As space and time change each of the mass eigenstates, $\nu_j$, all produced with the same energy, E, evolves acquiring a phase $-i(Et-p_jx)$, different for the 3 mass eigenstates due to their different masses (e.g. different momenta). When the neutrino is detected through a weak interaction, this quantum-mechanical mixture is projected again into a weak eigenstate, $|\nu_{\beta}>$, which can be different from $|\nu_{\alpha}>$. We say that the neutrino has oscillated from one flavor to another. Clearly the lepton family number is not conserved in this process. With some algebra one can compute the probability for the transition $|\nu_{\alpha}>$ to $|\nu_{\beta}>$:
\begin{eqnarray}
P (\nu_{\alpha} \rightarrow \nu_{\beta})  & = & \delta_{\alpha \beta} - 4 \sum_{i>j} 
	\Re (U^*_{\alpha i} U_{\beta i} U_{\alpha j} U^*_{\beta j}) \sin^2 [1.27 \Delta m^2_{ij} (L/E)] \nonumber \\
& +2 & \sum_{i>j} \Im (U^*_{\alpha i} U_{\beta i} U_{\alpha j} U^*_{\beta j}) \sin [2.54 \Delta m^2_{ij} (L/E)] 
\end{eqnarray}
\noindent In the above expression $\Delta m^2_{ij}\equiv m_i^2-m_j^2$ is measured in $(eV/c^2)^2$, $L$, the distance from the source to the detector, is measured in $km$, and $E$, the neutrino energy, is measured in $GeV$.

These equations simplify considerably if certain conditions are met. For example one of the mass splittings, $\Delta m$, can be very different from the other. If $E$ and $L$ in a given experiment are such that $1.27\Delta m^2(L/E)\approx \pi/2$ then only the corresponding term is relevant in the above expression. The resulting formulae are like those that one would obtain assuming that only two generations participate in the oscillation. The corresponding situation is called a ``quasi-two-neutrino oscillation". It can also be that only two mass states couple significantly to the flavor partner of the neutrino being studied. In that case the equations also become quasi-two-neutrino oscillations. Nature seems to be kind enough to have chosen these situations, the first in the case of atmospheric neutrinos the second in solar neutrinos.
\ \\

\subsection{Atmospheric neutrino oscillations}

The first clear signature for neutrino oscillations was reported by the Super-Kamikande collaboration at the time of the 1998 International Neutrino Conference in the study of atmospheric neutrinos \cite{SK1}. These are neutrinos produced by the interaction of the primary cosmic rays, high in the Earth atmosphere. The charged pions produced in the primary cosmic-ray collisions with the nuclei in the atmosphere decay into muons and muon-neutrinos (or antineutrinos). Most of the secondary negative (positive) muons also decay into an electron (positron), an electron antineutrino (neutrino) and a muon neutrino (antineutrino). The ratio of muon neutrinos to electron neutrinos arriving at the surface should then be about 2. Some of the muons do not decay before they reach the Earth surface so the ratio should be slightly larger than 2, an effect that grows with the energy of the primary muon (and thus of the resulting neutrinos). 
\ \\

The Super-Kamiokande detector is a huge, 50 kts, tank of water and it is located at 1 km depth at the Kamioka mine in western Japan \cite{sk-detector} (see below). Electron and muon neutrino interactions are detected when they interact through the charged current, giving an electron or muon crossing the detector or stopping in the water. The corresponding track produces a ring of Cherenkov light that is detected by the photomultipliers surrounding the detector volume. With this method one can measure the direction of the particle (electron or muon) which is strongly correlated with the direction of the incoming neutrino, and its energy \cite{sk-detector}. The ratio of electron to muon neutrinos reported by SuperKamiokande was closer to 1, instead of 2, but the most striking result was the zenith angle dependence of the high energy muon neutrinos shown in Figure 1 for the latest Super-Kamiokande data \cite{suzuki}, \cite{sk-dat}. The zenith angle is related to the distance traveled by the neutrinos before reaching the detector. Zenith angle equal 0 (cosine equal 1) means that the neutrinos enter the detector from the top, having been produced between 10 and 20 kilometers in the atmosphere, above the detector, while zenith angle equal $\pi$ implies neutrinos produced in the antipodes of the detector, at near 13,000 kilometers. From these plots one clearly sees that the expected number of neutrinos decreases with distance. Notice that the effect is present for muon neutrinos but not for electron neutrinos. 

\begin{figure}[h]
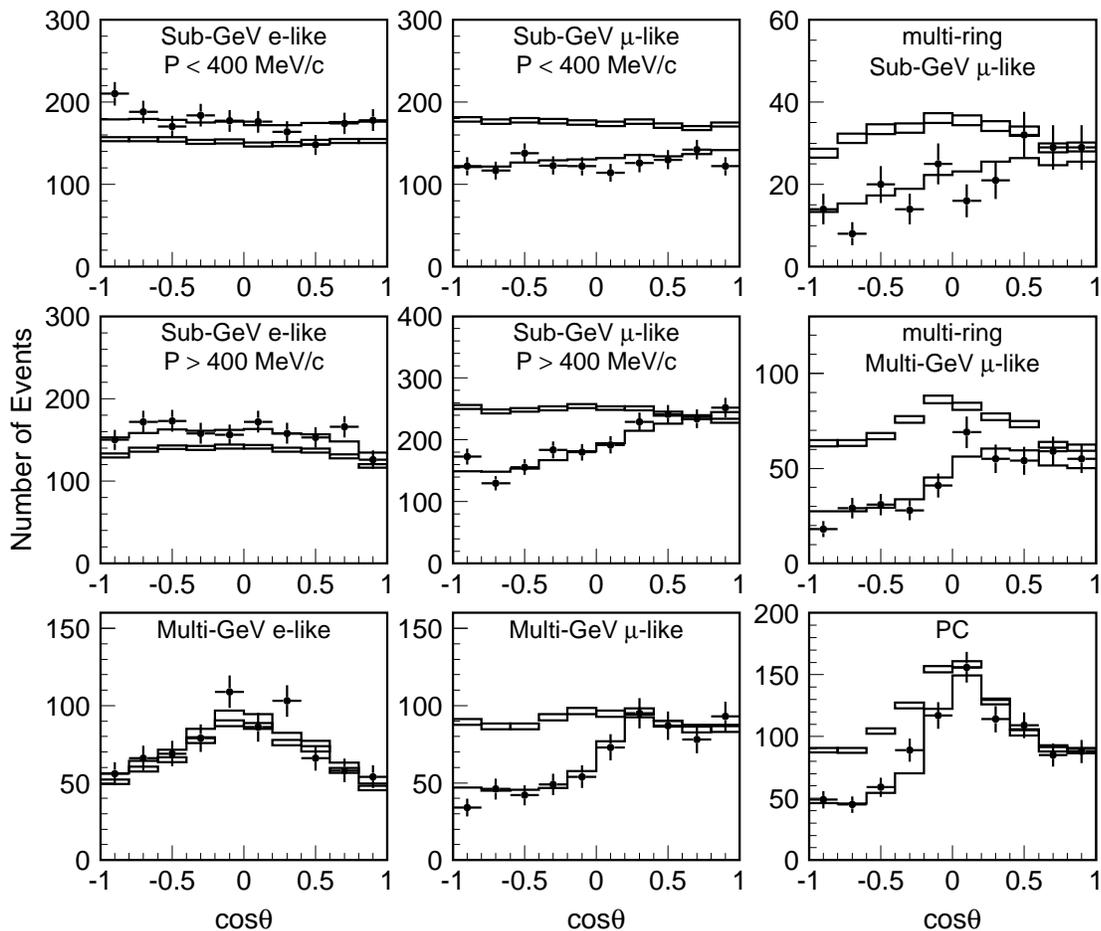

  \includegraphics[height=4.8in]{zenith_1a.epsi}
  \vspace{0.2in}
  \includegraphics[height=4.8in]{zenith_2a.epsi} 
 \caption{ \label{label} Number of electron and muon atmospheric neutrinos detected in Super-Kamiokande as a function of the zenith angle, for various types of events. cos$\theta=1$ corresponds to neutrinos coming from above (near) the detector while cos$\theta = - 1$ corresponds to neutrinos coming from below (far). The rectangles are the expected number of events in the absence of oscillations while the solid curve is the expectation for the best fit to the data including oscillations. The plots are from reference \cite{sk-dat}.}
\end{figure}

The results can be explained assuming that atmospheric muon neutrinos oscillate into tau neutrinos, which produce a tau in their interactions and cannot be easily identified in Super-Kamiokande. That the oscillation is predominantly into tau neutrinos and not electron neutrinos is derived from the fact that there is no anomaly on the electron neutrinos seen also in Super-Kamiokande. (The tau neutrinos interact producing a tau particle that decays promptly. These events either have too many Cherenkov rings from charged particles coming from the tau decay or cannot be distinguished from interactions of electron or muon neutrinos. An attempt to extract a tau neutrino signal is described in \cite{suzuki}). The oscillation parameters are such that the effect is more pronounced for those coming from long distances, order the Earth diameter, for the energies typical of atmospheric neutrinos. In fact, atmospheric neutrino oscillations can be described by the two-neutrino approximation assuming maximal mixing and one mass difference, called $\Delta m^2_{atm}$, with a value which until recently was mostly determined by the results of this experiment (see below).

\subsection{Solar neutrino oscillations}

If the first clear signature of neutrino oscillations appeared in atmospheric neutrinos in 1998, the first hint for oscillations came from an experiment that started in the mid sixties by R. Davis Jr. and collaborators with the aim of measuring the neutrinos originating in the Sun. This experiment reported a deficit \cite{davis1} in the solar neutrino flux with respect to the expectations from a solar model calculation of J. Bahcall and collaborators, published in the same issue of the Physical Review Letters in 1968 \cite{bahcall1}. As for any other star, electron neutrinos are produced in the Sun due to the fusion reactions taking place deep in the star interior, as first written down by Hans Bethe in the thirties \cite{bethe}. Most of the neutrinos are produced in the fusion of two Hydrogen nuclei (two protons) into Helium, giving neutrinos with an energy lower than 0.42 MeV, which is below the detection threshold of most experiments. But there are other neutrino producing reactions taking place in the Sun, giving neutrinos with a rich spectrum of energies. After the pioneer experiment many others have followed, using different techniques, sensible to different parts of the solar neutrino spectrum. In particular there are neutrinos produced in the beta decay of $^8B$ which have and end point energy of 15 MeV and are the best measured (again the reader is referred to the lecture by D. Wark in this conference). 

Almost all the solar neutrino experiments reported a measured electron neutrino flux which was substantially below the expectations of the solar model, as low as 30\% depending on the energy \cite{solar-review}. The interpretation was that the solar electron neutrinos oscillate into muon or tau neutrinos which are below threshold for the charged current interaction, in which a muon or a tau, respectively, has to be produced. Obviously the way to avoid this detection problem is to design an experiment able to detect the neutral current interaction, which should take place for the three neutrino species. This is precisely what the SNO experiment has achieved. The SNO detector \cite{SNO} consists on a tank of 1 kt (kiloton) of heavy water, $D_2O$, shielded by about 7.5 kt of ultra-pure ordinary water. Electron neutrinos can interact in SNO by the charged current (CC) reaction that transforms the neutron of the deuterium nucleus into a proton
\begin{equation}
\nu_e + d \rightarrow p + p + e^-
\end{equation} 
\noindent The electron is observed by the Cherenkov light and its direction is strongly correlated with that of the neutrino. But what is unique to SNO is that the three types of neutrinos can interact with the deuterium via the neutral current reaction (NC)
\begin{equation}
\nu_x + d \rightarrow \nu_x + n + p 
\end{equation} 
\noindent The threshold energy for this reaction is 2.22 MeV, the binding energy of the deuterium nucleus. This reaction is detected through the observation of the neutron. In the first phase of SNO the neutron was detected when it was captured by the deuteron emitting a 6.25 MeV $\gamma$-ray. In a second phase of the experiments 2 tons of a salt, $NaCl$, were added to the heavy water and the neutron was detected following its capture by the $^{35}Cl$ nucleus, which produces a $\gamma$ cascade. The efficiency of detecting the neutron is three times higher with the salt added to the water. To be able to detect above background the single electron of the CC events, or the light due to the single neutron capture in the NC events, requires an extremely high purity of the detector elements. A third reaction is the elastic scattering (ES) of neutrinos with electrons (this is the reaction also measured in SuperKamiokande)
\begin{equation}
\nu_x + e^- \rightarrow \nu_x + e^- 
\end{equation} 
\noindent This reaction occurs in principle for the three neutrino types (through the neutral current) but the cross section is higher for electron neutrinos, where it can also proceed via the charged current. SNO has been able to measure the three reactions and from them infer a flux of neutrinos from the Sun. The solar flux inferred from these reactions separately, from the data of the second phase and for an effective electron kinetic energy threshold of 5 MeV in the CC and ES reactions, is \cite{poon}:
\begin{eqnarray}
\Phi_{CC} & = & 1.68 \pm 0.06 (stat.)^{+0.08}_{-0.09} (sys.) \times 10^6 \ cm^{-2}s^{-1} \nonumber \\
\Phi_{ES} & = & 2.35 \pm 0.22 (stat.)^{+0.15}_{-0.15} (sys.) \times 10^6 \ cm^{-2}s^{-1} \nonumber \\ 
\Phi_{NC} & = & 4.94 \pm 0.21 (stat.)^{+0.38}_{-0.34} (sys.) \times 10^6 \ cm^{-2}s^{-1} 
\end{eqnarray}
The excess of the NC flux over the other fluxes is a clear indication that neutrinos from the Sun change flavor. Furthermore the total flux from NC is in very good agreement with the total $^8B$ flux of $5.05^{+1.01}_{-0.81} \times 10^6 \ cm^{-2}s^{-1}$ predicted by the SSM \cite{bahcall}.  The results are illustrated in Figure 2, which shows the flux of muon plus tau neutrinos versus the flux of electron neutrinos which are deduced from the SNO measurements \cite{sno-salt1}, \cite{sno-salt2}. As it can be seen the CC reaction gives only the electron-neutrino flux, while the other two reactions, (NC) and (ES), give two different linear combinations of $\Phi_e$ and $\Phi_{\mu\tau}$. The data on ES from Super-Kamiokande is also included in the Figure. The bands intersect at one point, giving a consistent solution. The band bounded by the dotted lines is the prediction of the SSM. A long standing problem in physics, the deficit of solar neutrinos with respect to the SSM, is now finally solved.

\begin{figure}[h]
\begin{minipage}{18pc}
\includegraphics[width=18pc]{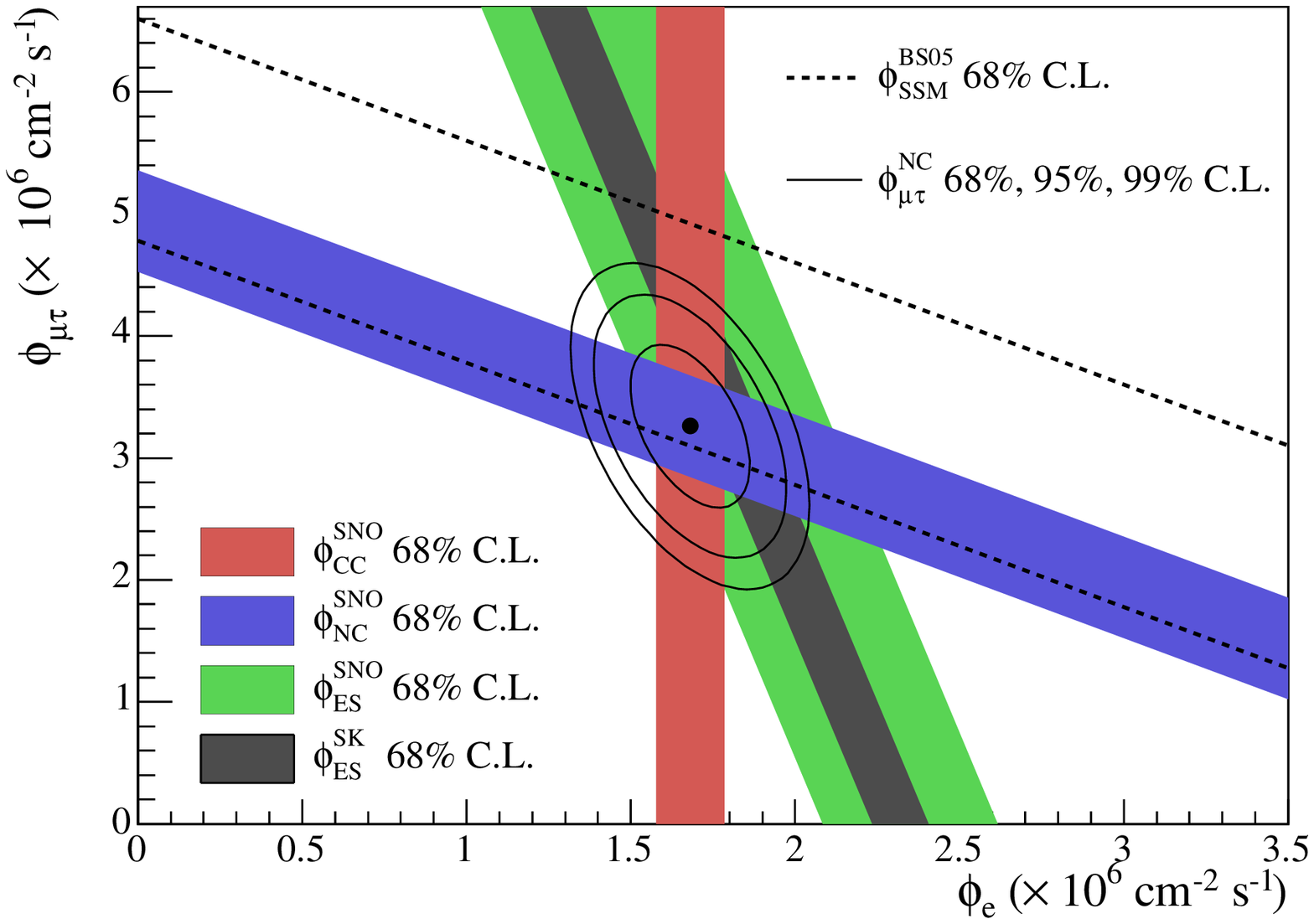}
\caption{\label{label} The solar neutrino fluxes inferred from the SNO and Super-Kamiokande measurements. The band bounded by the dotter lines is the prediction of the standard solar model \cite{bahcall2} while the parallel solid band is that extracted from the measurement of the neutral current reaction (see the text for an explanation). The figure is from \cite{sno-salt2}.}
\end{minipage}\hspace{2pc}%
\begin{minipage}{18pc}
\includegraphics[width=18pc]{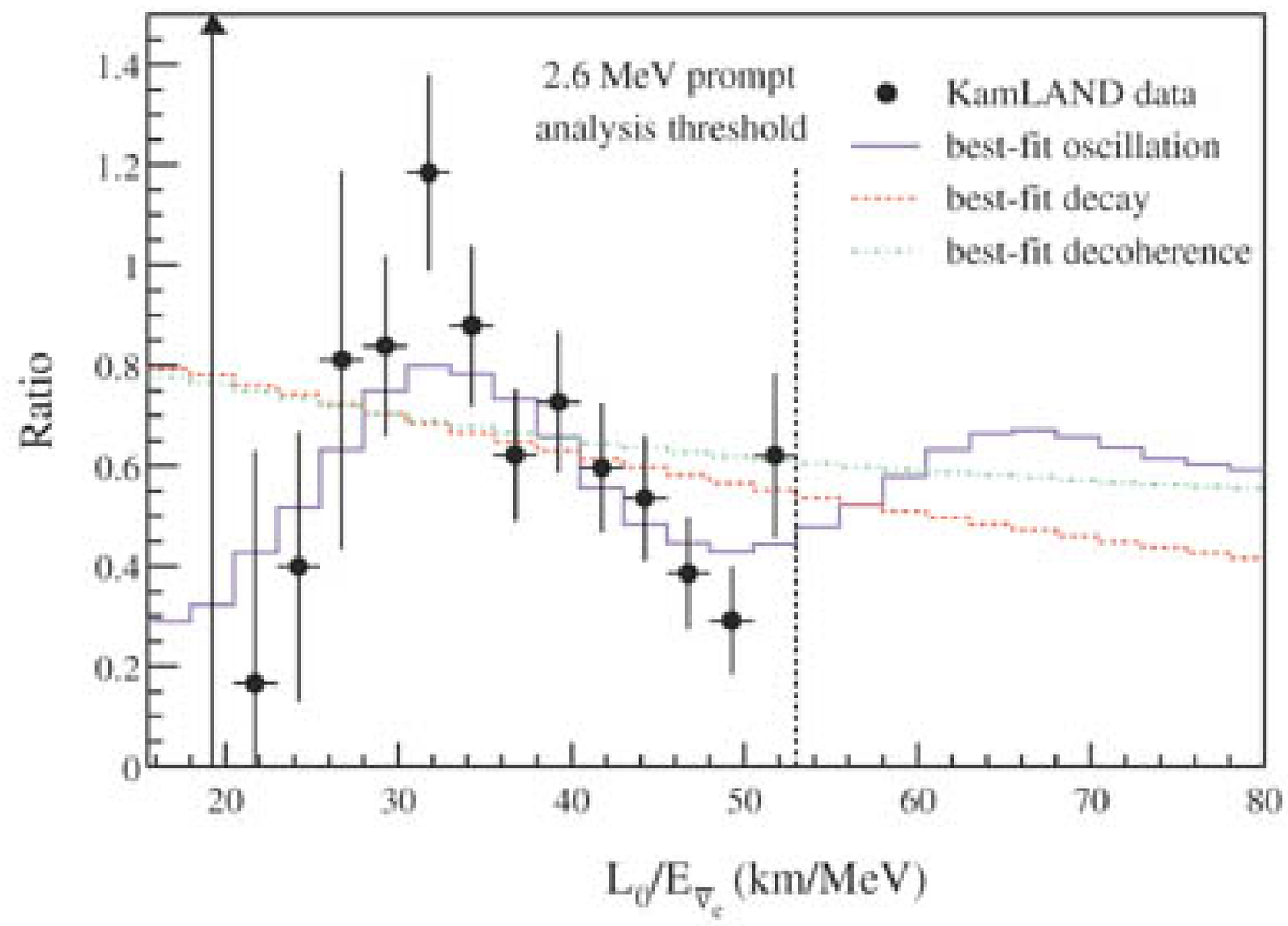}
\caption{\label{label} Ratio of the observed $\bar{\nu_e}$ spectrum to the expectation for non-oscillation versus $L_0/E$ obtained by the KAMLAND collaboration. In the analysis $L_0$ is fixed to 180 km. The solid curve is the expectation for oscillations while the other two lines are the expectation for other models of neutrino disappearance. The plot is from \cite{kamland}.}
\end{minipage}
\end{figure}

Another recent experiment, KAMLAND \cite{kamland-det} has confirmed the solar neutrino oscillations. This experiment detects reactor antineutrinos produced in nuclear reactors in Japan, in one kiloton of liquid scintillator. The typical distances from source to detector (average 180 km) and energies of the antineutrinos (about 3 MeV average) make the experiment sensitive to the solar oscillation parameters, in this case in vacuum (matter effects can be neglected over these distances). The experiment detected $258$ antineutrino interactions while $365.2 \pm 23.7$ were expected, with a background of $17.8\pm 7.3$ \cite{poon}, \cite{kamland}. The difference is inconsistent with a square distance decrease of the flux and it is attributed to oscillations. Furthermore the $L/E$ distribution of these antineutrinos, shown in Figure 3, also shows the characteristic distortion expected from oscillations. The KAMLAND results improve significantly the measurement of the value of $\Delta m^2_{12}$ with respect to the solar experiments alone (see below).

\subsection{Matter enhanced oscillations}

The results of solar neutrinos are in agreement with the solar model but they also indicate that the neutrinos born as $\nu_e$ neutrinos in the Sun change their nature on their way to the detector on Earth. A full analysis of the problem requires the study of neutrino propagation in a dense medium and its interplay with oscillations. When the neutrinos propagate in a dense medium, such as the Sun core or the Earth interior, they can coherently forward-scatter (refer to the lectures of A. Smirnov and J. Steinberger in this workshop) with electrons, neutrons and protons in the medium. All neutrino types can forward interact through the neutral current but electron neutrinos can also forward-interact with electrons in the medium via a charged current. Coherent forward scattering gives rise to an extra phase that affects neutrino propagation in matter. This phase is the same for the three neutrino species except for that extra piece in the electron neutrinos due to the charged current interaction with the electrons in the medium. This extra phase, which has nothing to do with mixing and mass differences, has to be added to that due to mixing, and this can have dramatic consequences in the oscillation. The extra phase is proportional to the electron density in the medium, to the Fermi constant $G_F$ and to the energy of the neutrino. The oscillation formulae are similar to those in vacuum, with the replacement of the mass differences and mixing angles by effective quantities which depend on the medium \cite{kayser}. The net effect is that the oscillation probabilities can be greatly enhanced with respect to those in vacuum, an effect known as the Mikheyev-Smirnov-Wolfenstein (MSW) effect \cite{MSW}.

It can be shown that in the case of solar neutrinos the matter effect is important for solar densities and for neutrino energies above a few MeV, certainly for those coming from the $^8B$ reaction. The neutrinos born as $\nu_e$ in the Sun propagate to its surface, where they arrive as the pure mass state denoted by $\nu_2$, because of matter-enhanced transition. From there they propagate to the Earth (without oscillating, since they are in a pure mass state) until they interact. The probability of producing an electron neutrino, that is, the electron neutrino survival probability, is a direct measure of the electron neutrino component of $\nu_2$. This is essentially the ratio of CC/NC measured by SNO. One of the goals of the SNO in the future is to improve this measurement \cite{poon}.

The above interpretation of the data in terms of oscillations with three neutrino mass states, and indeed most of the analysis extracting oscillation parameters from neutrino experiments, makes use of another two results. One is the result of the Chooz \cite{chooz} reactor experiment, in which no disappearance of electron antineutrinos (produced in a nuclear reactor) was seen over a short baseline of 1 km. This result, together with the mass difference for atmospheric neutrinos, implies that the $\nu_3$ neutrino has a negligible component of electron neutrinos (see below). The other assumption is that the LNSD \cite{LNSD} result is not valid, otherwise the results from oscillations cannot be accommodated in a scheme with just 3 mass states.

\section{Oscillation parameters}

The flavor content of the mass states and a summary of what the different experiments contribute to the knowledge of the oscillation parameters is shown in Fig. 4, taken from the article of B. Kayser \cite{kayser}. As it can be seen a lot is known already about lepton mixing and mass differences, however we do not know the mass hierarchy, that is, we do not know if the mass of the state $\nu_3$ is larger or smaller than the masses of the states $\nu_1$ and $\nu_2$. This is because 
$\Delta m^2_{32}\equiv m_3^2-m_2^2 \equiv \Delta m^2_{atm}$ is determined from atmospheric oscillations which are not sensitive to the square of the mass differences (the oscillation probability depends on the $sin^2$ of this quantity). For $\nu_1$ and $\nu_2$ however we know the sign of the mass difference, since the result is based on the solar oscillations, 
$\Delta m^2_{21}\equiv m_2^2-m_1^2 \equiv \Delta m^2_{sol}$ 
for which matter effects, which depend on the sign of this quantity, are important.

\begin{figure}[h]
\begin{minipage}{26pc}
\includegraphics[width=26pc]{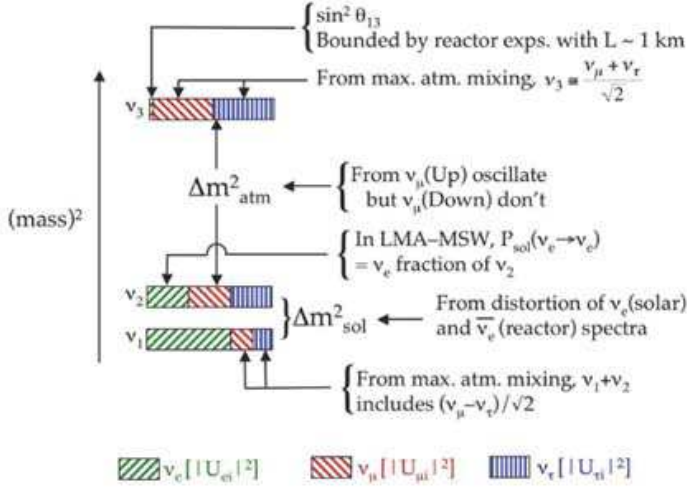}
\caption{\label{label}The neutrino mass-square spectrum from present neutrino oscillation experiments. The figure has been taken from \cite{kayser}.}
\end{minipage}
\end{figure}

When extracting oscillation parameters the mixing matrix is conveniently parameterized in the following way:
\begin{equation}
U =  \left ( \begin{array}{ccc} 1 & 0 & 0 \\ 0 & c_{23} & s_{23} \\
	0 & -s_{23} & c_{23}\end{array} \right ) \  
     \left ( \begin{array}{ccc} c_{13}& 0 & s_{13}e^{ - i\delta} \\ 
      0 & 1 & 0 \\ - s_{13}e^{i\delta} & 0 & c_{13} \end{array} \right ) \
     \left ( \begin{array}{ccc} c_{12} & s_{12} & 0 \\ -s_{12} & c_{12} & 0 \\
	0 & 0 & 1 \end{array} \right) \  
     \left ( \begin{array}{ccc} e^{i\alpha_1/2} & 0 & 0 \\ 
     0 & e^{i\alpha_2/2} & 0 \\ 0 & 0 & 1 \end{array} \right ) 
\end{equation}
\noindent where $c_{ij}=cos\theta_{ij}$, $s_{ij}=sin\theta_{ij}$. The two phases $\alpha_1$ and $\alpha_2$ are only present if the neutrino is a Majorana particle \cite{majorana} and they do not affect oscillations or the interpretation of present neutrino results. The first matrix governs essentially the oscillation of atmospheric neutrinos while the third governs the oscillation of solar neutrinos. The angle $\theta_{23}$ is essentially the angle extracted from treating atmospheric oscillations as a two-neutrino oscillations, and the angle $\theta_{12}$ is essentially the angle extracted from solar experiments treating the solar oscillations as two-neutrino oscillations.

The extraction of oscillation parameters from global fits to all the available data has been studied by several authors, both in the context of three neutrino mass states or, including the LSND results, four mass states. Here we only show data for three mass states. The results have been taken from a recent review  \cite{goswami}. Figure 5 shows the allowed area in the $\Delta m^2_{21}$ versus $sin^2\theta_{12}$ plane, including the solar and KAMLAND results. The plot is taken from the review of S. Goswami \cite{goswami}. Figure 6 is a similar plot of $\Delta m^2_{atm}$ versus $sin^2\theta_{atm}$, obtained from atmospheric and K2K (see below) results \cite{goswami}, \cite{maltoni}.

In the few years since the first atmospheric results, the progress in the measurement of the leptonic mixing matrix has been enormous. What we do not yet is the value $\theta_{13}$ (the electron component of $\nu_3$), for which only the upper limit $sin^2 \theta_{13}<0.04$ (or $sin^2 2\theta_{13}<0.15$) is known, mainly because of the negative result of the Chooz experiment \cite{chooz}. The other completely unknown parameter is the phase $\delta$. A non-zero $\delta$ value will in general lead to CP violation , that is, to a difference in the oscillation probability of $\nu_{\alpha} \rightarrow \nu_{\beta}$ with respect to $\bar{\nu}_{\alpha} \rightarrow \bar{\nu}_{\beta}$. It is clear that, because of the way $\theta_{13}$ and $\delta$ enter in the mixing matrix, the ability to measure $\delta$ in an experiment depends on the value of $\theta_{13}$. And we do not know yet the mass hierarchy, that is, whether $m_3$ is larger or smaller than $m_1$ and $m_2$. One of the main goals of the future accelerator experiments is to measure these quantities.

\begin{figure}[h]
\begin{minipage}{30pc}
\includegraphics[width=30pc]{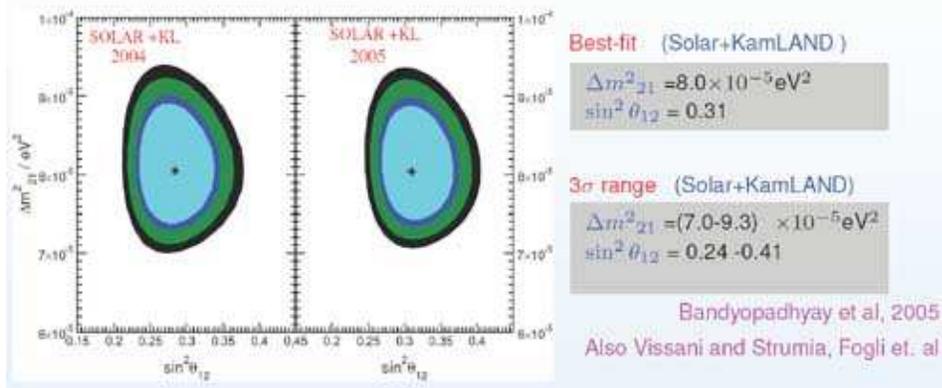}
\caption{\label{label}The allowed region of the $\Delta m^2_{21}$ versus $sin^2\theta_{12}$ plane from solar neutrino and KAMLAND experiments \cite{goswami}.}
\end{minipage} 
\end{figure}

\begin{figure}[h]
\begin{minipage}{30pc}
\includegraphics[width=30pc]{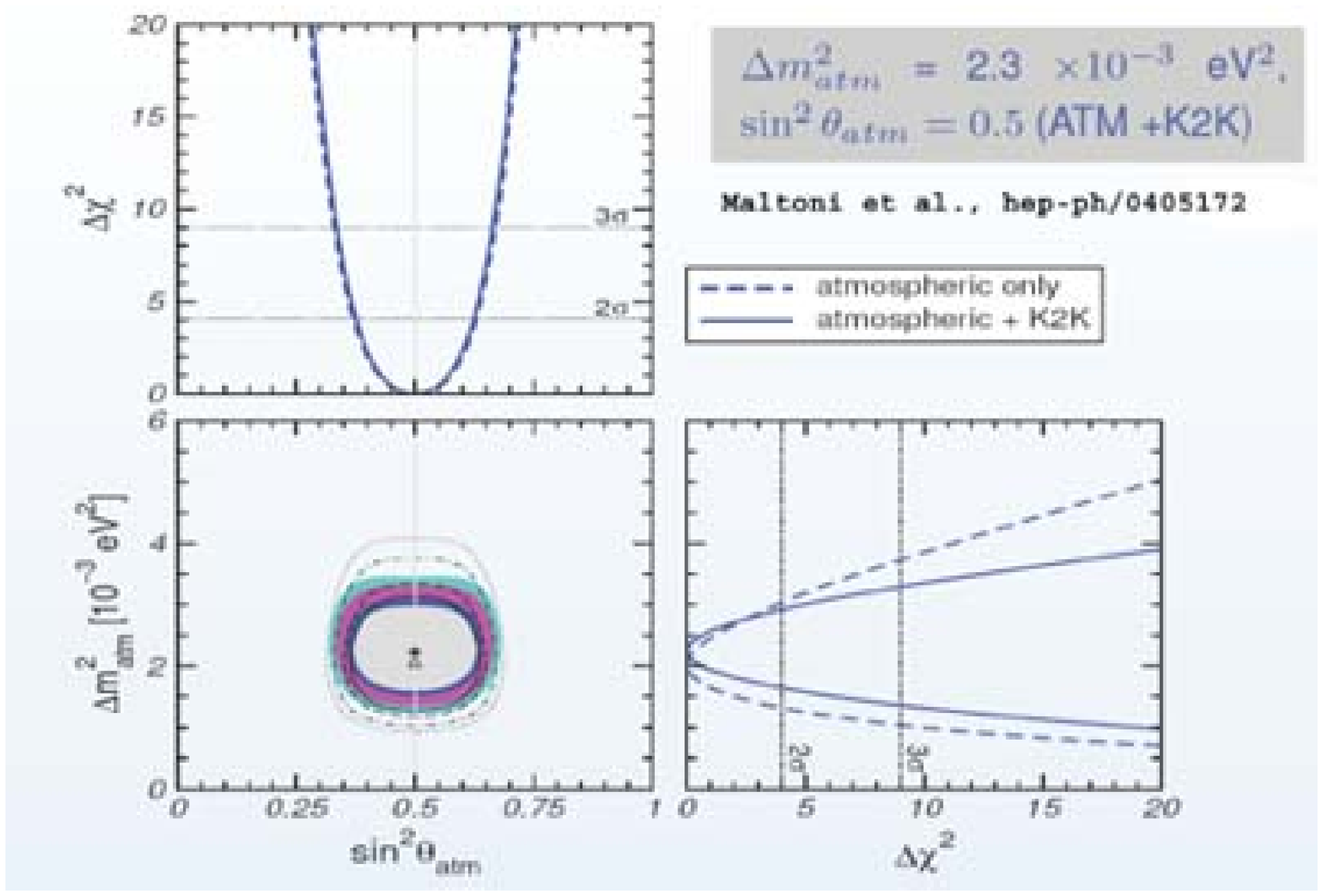}
\caption{\label{label} The allowed region of the $\Delta m^2_{atm}$ versus $sin^2\theta_{atm}$ plane from atmospheric and K2K neutrino experiments \cite{goswami}, \cite{maltoni}.}
\end{minipage} 
\end{figure}

\section{Accelerator long baseline experiments}

As we have seen neutrino oscillations in atmospheric and solar neutrinos involve large distances, except for the positive result of the LNSD experiment, where evidence was found for muon antineutrino oscillation into electron antineutrinos \cite{LNSD}. The baseline for this experiment was only $30m$ and the energy of the antineutrinos was below 55 MeV, since most of them come from positive muons decaying at rest. The LNSD results are now being checked by an experiment designed for that purpose: the MiniBooNE experiment (BooNE stands for Booster Neutrino Experiment) now taking data at FNAL \cite{miniboone}. This experiment has a longer baseline than LNSD (about $0.5$ km) and also a larger neutrino energy (average of 0.8 GeV). The experiment can run with neutrinos or antineutrinos, by changing the polarity of the focusing horn. The results of this experiment are eagerly expected.

The first long baseline experiment to have come into operation was K2K, which took data from June of 1999 to November of 2004 \cite{ahn}. The primary goal of this experiment was to study the oscillation parameters that govern atmospheric neutrino oscillations, by looking at the disappearance of muon neutrinos over a baseline of 250 kilometers. Another experiment with the same primary goal is MINOS, presently taking data. In the near future, the CNGS (CERN to Gran Sasso) project will also start, in this case with the goal to verify explicitly that the atmospheric oscillation is indeed the conversion of $\nu_{\mu}$ to $\nu_{\tau}$. These experiments are described in this section.

\subsection{The K2K Experiment}
The beam for K2K (KEK to Kamioka) was produced by focusing positive pions produced in collisions of 12 GeV protons from the KEK-PS accelerator. The focusing is achieved by a system of two magnetic horns, located next to the proton target. Following the horns the pions enter a 200m long decay pipe where they are allowed to decay into muons and neutrinos. The pipe is tilted downwards by $1^0$, in such a way that the beam of neutrinos, produced in in-flight $\pi^+$ decays, points to the Super-Kamiokande detector, located at a distance of 250 km (Figure 7).

\begin{figure}[h]
\begin{minipage}{30pc}
\includegraphics[width=30pc]{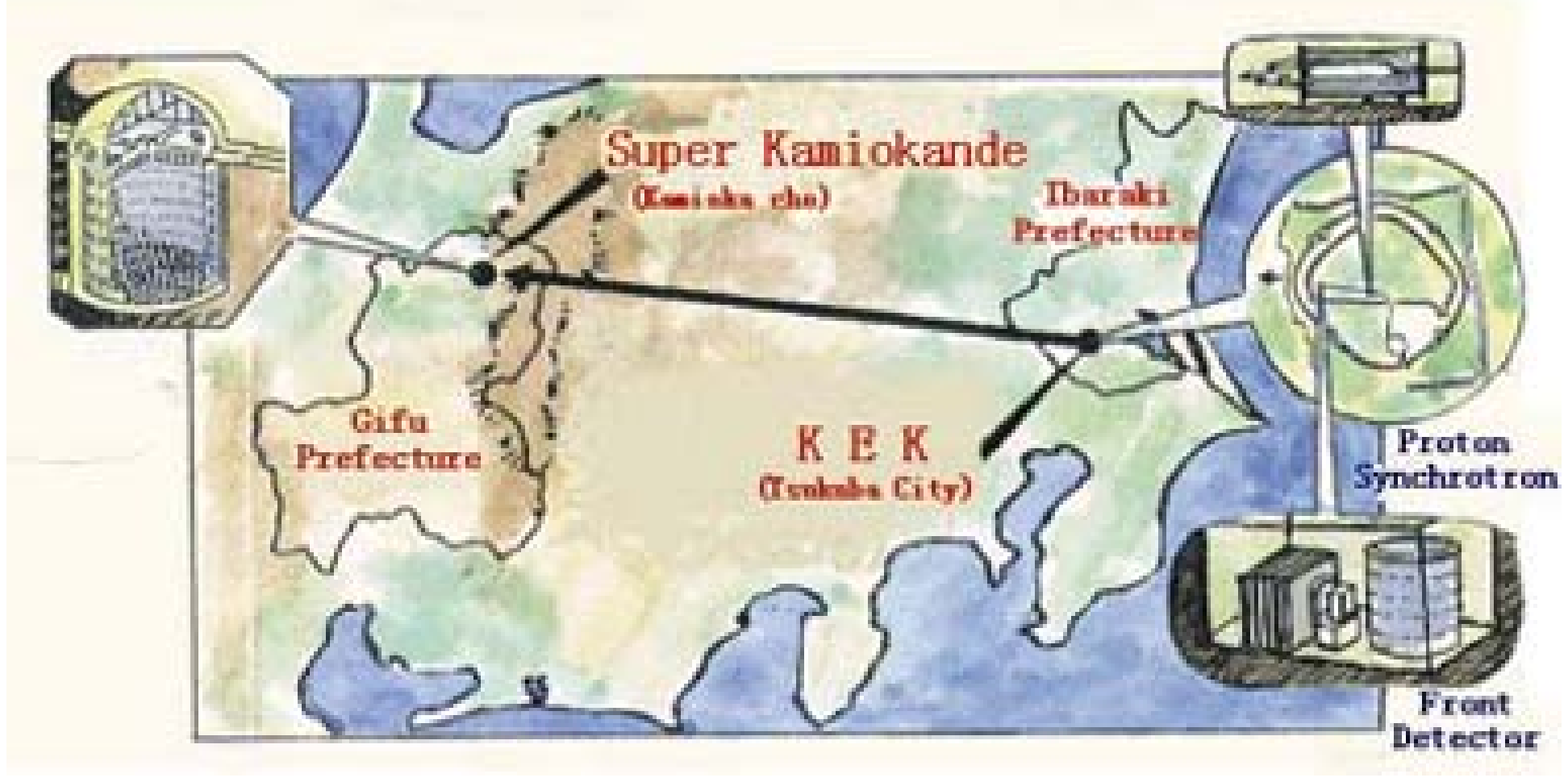}
\caption{\label{label}Scheme of the K2K experiment. The neutrino beam produced by the KEK-PS accelerator is directed towards the Super-Kamiokande detector, located at a distance of 250 km.}
\end{minipage}
\end{figure}

That the beam points in the correct direction is ensured by a Muon Monitor, located after an iron shielding at the end of the decay pipe. The muon monitor measures the profile of the muons that reach that detector, which have an energy above 5.5 GeV. Since the muons and the neutrinos come from the same parent pions they follow the same direction, which is monitored all the time and known with a precision of better than $1mrad$. A $3mrad$ change on the direction induces a 1\% change in the flux and neutrino energy spectrum in SK, which are therefore stable at much better than 1\%. The beam stability is also checked directly by studying the spectrum and direction of muons produced in neutrino interactions in the near detector, namely in the Muon Range Detector (MRD) described below.
\ \\

The oscillation in K2K is measured by comparing the expected number and energy distribution of muon neutrinos reaching SK with the corresponding quantities that are actually measured. The expected number of events in SK is inferred from the neutrino interactions detected in a set of near detectors (ND), located approximately $300m$ downstream of the proton target in KEK (see Figure 8). 

\begin{figure}[h]
\begin{minipage}{30pc}
\includegraphics[width=30pc]{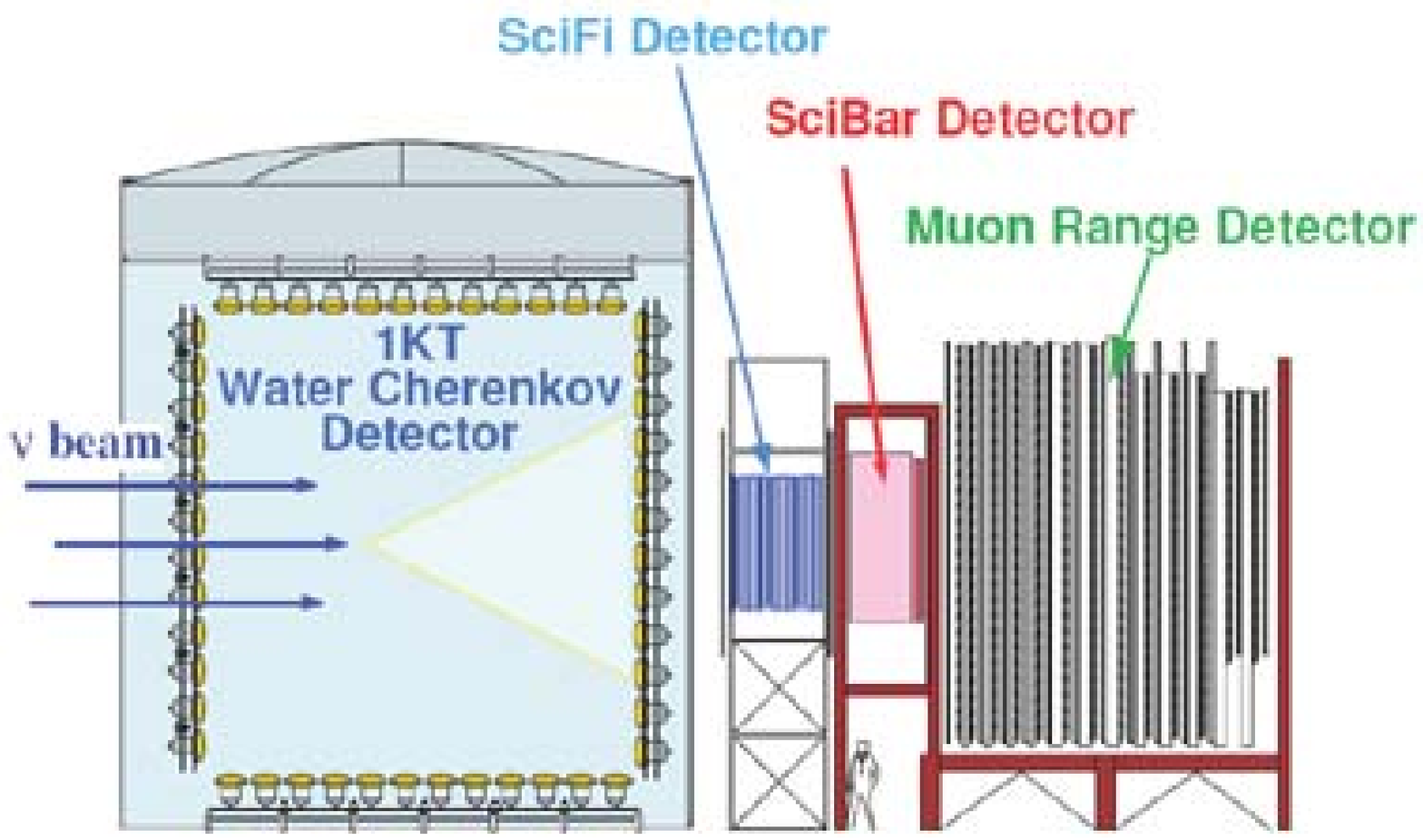}
\caption{\label{label} Scheme of the K2K near detector system. See text for an explanation of the detector components.}
\end{minipage}
\end{figure}
\ \\

The ND set comprises a 1 kt water Cherenkov (1KT)which is a scaled down version of the SK detector, and a fine-grained detector system (FGD). The 1 kt cylindrical water tank is optically separated into an inner detector, viewed by 680 photomultiplier tubes (PMT) of 50cm diameter giving a photocathode coverage of 40\%, and an outer detector, viewed by 68 PMT of 20cm diameter, facing outward. The FGD consists of a scintillating-fiber/water tracker (SciFi), a lead-glass calorimeter (LG) used in the first part of the running period (K2K-I), replaced by a totally active fine-segmented scintillator tracker (SciBar) during the second part of the running (K2K-II), and a muon-range detector (MRD). The Super-Kamiokande detector is the far detector and it consists of a 50 kt cylindrical water tank, with a diameter of 39m and a height of 41m\cite{sk-detector}. The tank is divided into two optically separated regions. The inner region has a diameter of 33.8m and a height of 36.2m, it contains 32 kts of water and it is viewed by 11,146 PMT of 50cm diameter, corresponding to a 40\% photocathode coverage during phase I, and by 5,182 PMT during phase II (after an accident in November of 2002 when approximately half of the PMTs were broken). The outer region is viewed by 1,885 8-inch PMTs. A fiducial volume is defined in SK as a cylinder whose surface is 2m away from the inner detector surface, providing a fiducial mass of 22 kts.

The method for the oscillation analysis is the following \cite{k2k-long}. First the neutrino flux and spectrum are measured in the near detector before oscillation. The total flux is inferred from the event rate in the 1KT detector, while the spectrum is computed independently from the information in the 1KT, SciFi, LG and SciBar detectors and from their information combined. Then an extrapolation of these quantities to the far detector is made, assuming no oscillation. This extrapolation is calculated by Monte Carlo, supplemented with several measured ingredients. Because of the finite size of the detectors and of the diameter of the decay pipe the far to near ratio of the flux (F/N) does not follow an inverse distance-squared law, but it is a function of the energy. The calculation gives, for each neutrino energy, the ratio
\begin{eqnarray} R^{F/N} = \frac{\Phi^{SK}(E_{\nu})}{\Phi^{ND}(E_{\nu})} \end{eqnarray}

The calculation can be carried out precisely if one knows the spectrum and angle of the pions produced in the primary proton interactions with the target. For this experiment the pion production was taken from a parameterization of pion cross-sections and from the direct measurement of pion production in the HARP \cite{harp} experiment, for protons of the same energy than that of the KEK-PS, colliding with a target identical to that of K2K. This information was also checked directly by measuring the pions downstream of the horns by means of a detector, named PIMON, which was occasionally put into the beam. Once the far to near ratio is know one can compute the expected number of events in SK and their spectrum assuming no oscillation, and compare them with what is actually measured.

The neutrino beam is wide band, with an average energy of the neutrinos in the near detector of about 1.3 GeV. In this energy region the highest cross section is that of the quasi-elastic interaction
\begin{eqnarray} \nu_{\mu} + n \rightarrow \mu^- + p \end{eqnarray}
\noindent For this reaction the energy of the neutrino can be calculated if the energy and angle of the muon are known, assuming that the neutron is at rest, that is, neglecting the Fermi momentum in the nuclear target. This energy is given by the expression
\begin{eqnarray} E_{\nu}=\frac{m_N E_{\mu}-m^2_{\mu}/2}{m_N - E_{\mu}+ P_{\mu} cos\theta_{\mu}} \end{eqnarray}
\noindent where $m_N$, $E_{\mu}$, $P_{\mu}$ and $\theta_{\mu}$ are the nucleon mass, the muon energy, the muon momentum and the muon scattering angle respectively. When measuring the neutrino spectrum in the different detectors, samples of quasi-elastic enriched events are selected. The reader is referred to \cite{k2k-long} for the details.

To select events in SK a timing cut is applied around the expected arrival of the beam, calculated with the use of the GPS global positioning system which provides synchronization with an accuracy better than 200ns. The beam spill at the KEK-PS is 1.1 $\mu s$ and contains 9 bunches with 125 $ns$ time interval between them. This time structure is clearly observed in the selected events at SK. By requiring that the events occur in a time window of 1.2 $\mu s$, that they have an energy deposition of more than 30 MeV and that they are fully contained in the fiducial volume of 22 KT, a total of 112 events were selected in SK \cite{suzuki}, \cite{ahn}, \cite{k2k-long}, \cite{aliu}. The extrapolation from the samples measured in the ND in absence of oscillation was $155.9\pm0.3^{+13.6}_{-15.6}$. Of the 112 events 67 were classified as single ring (58 muon-like and 9 electron-like) and 45 as multi-ring. The neutrino energy distribution of the 58 muon-like 1-ring events, computed assuming that they are quasi-elastic, is shown in Figure 9. The lines, normalized to the number observed are for the no-oscillation hypothesis (undistorted spectrum) and for oscillation with the best fit parameters. The allowed parameter region for K2K is shown in Figure 10. The best fit value in the physical region is $(\Delta m^2,sin^22\theta)=(2.8\times10^{-3}eV^2,1.0)$, which is consistent with the atmospheric neutrino oscillation, thus confirming the latter with artificially produced neutrinos.

\begin{figure}[h]
\begin{minipage}{18pc}
\includegraphics[width=18pc]{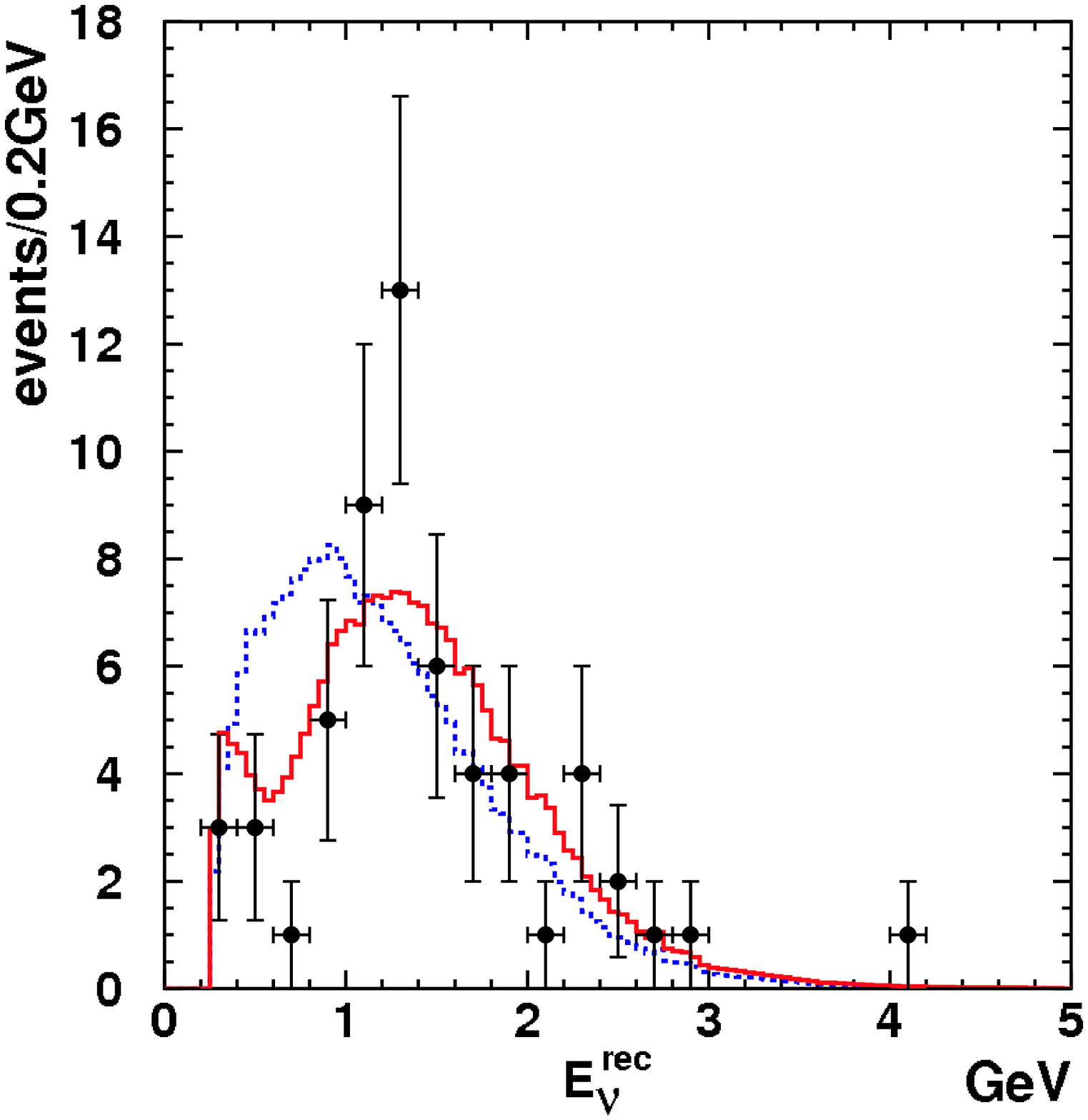}
\caption{\label{label} The spectrum of 1-ring muon-like events measured in Super-Kamikande.}
\end{minipage}\hspace{2pc}%
\begin{minipage}{18pc}
\includegraphics[width=18pc]{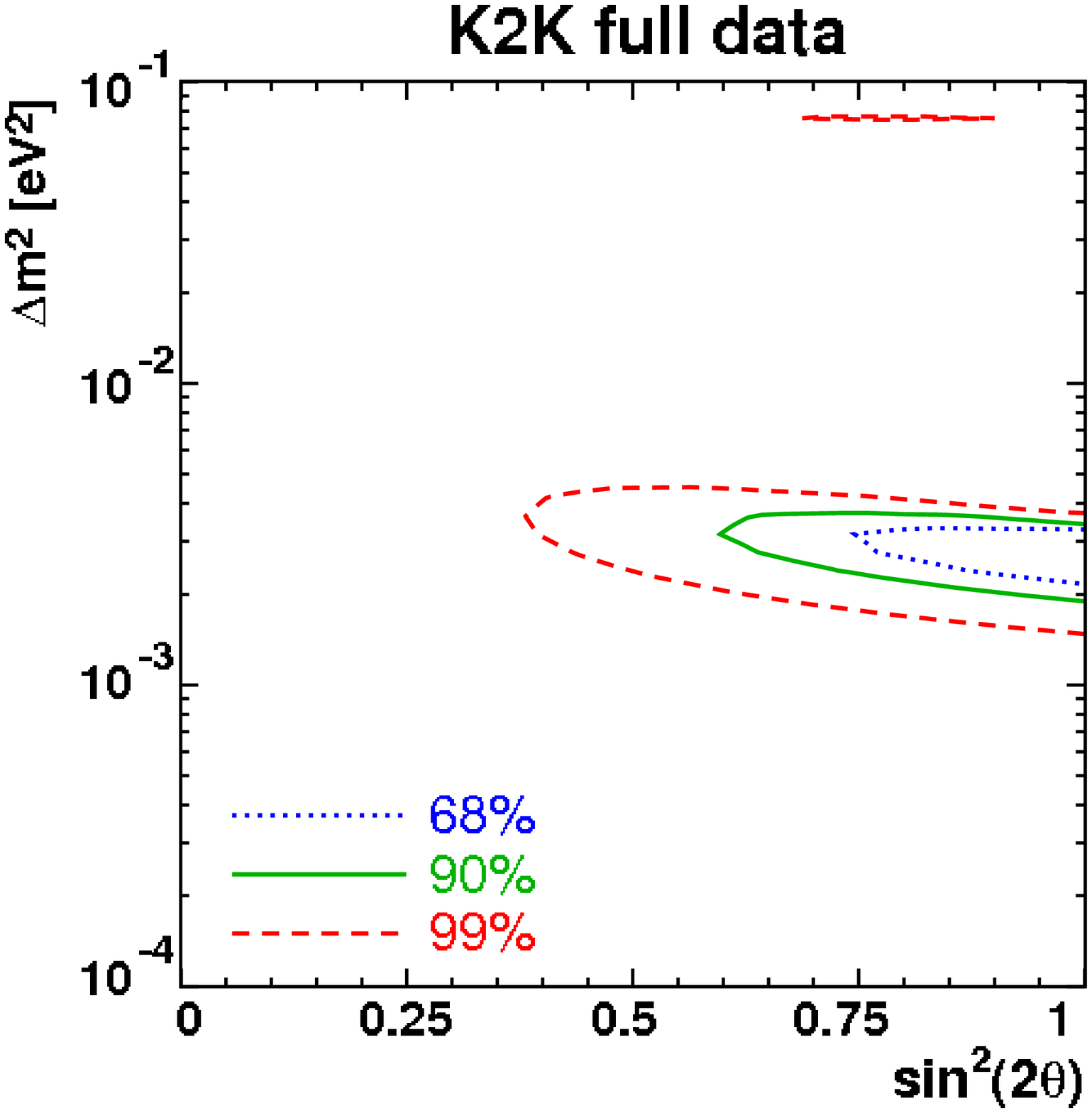}
\caption{\label{label} The allowed regions in the $\Delta m^2, sin^2 2\theta$ plane for three different CL \cite{k2k-long}.}
\end{minipage}
\end{figure}
\ \\

K2K has also given a limit for $\nu_{\mu}$ to $\nu_e$ oscillation over the baseline and energies of the experiment \cite{suzuki}, \cite{k2k-nue}. This analysis starts from the 9 e-like events found in the above analysis. Tighter electron identification cuts are applied to these events, designed to improve the purity of the electron sample, in particular to reject $\mu \rightarrow e$ decays. These tighter cuts retain 5 events. Of these remaining events most come from $\pi^0$ decays in which one of the $\gamma$ from the decay was not detected and the other converted producing e-like tracks. Looking for second rings and rejecting events if the reconstructed mass was consistent with a $\pi^0$ removes all but 1 event. The expected background is 1.3 events from $\nu_{\mu}$ interactions and 0.4 events from intrinsic $\nu_e$ component in the beam. The 1 event is therefore attributed to background, giving a 90\% CL upper limit for $\theta_{\mu e}$ of $sin^2 2 \theta_{\mu e}<0.13$ at 
$\Delta m^2_{\mu e}=2.8\times10^{-3} eV^2$. These parameters are those coming from treating the oscillation in the two neutrino approximation. To a very good approximation $\Delta m^2_{\mu e}\approx \Delta m^2_{23}$ and $sin^2 2\theta_{\mu e}\approx sin^2\theta_{23}\sin^2\theta_{13}\approx\frac{1}{2}sin^22\theta_{13}$.

\subsection{The MINOS Experiment}
The MINOS (Main Injector Neutrino Oscillation Search) experiment is a long baseline experiment presently taking data \cite{thomson}. The main objective of the experiment is to study the oscillation parameters of atmospheric neutrino oscillations by looking for $\nu_{\mu}$ disappearance in an accelerator neutrino beam, in particular to improve the measurement of $\Delta m^2_{23}$.

\begin{figure}[h]
\begin{minipage}{18pc}
\includegraphics[width=18pc]{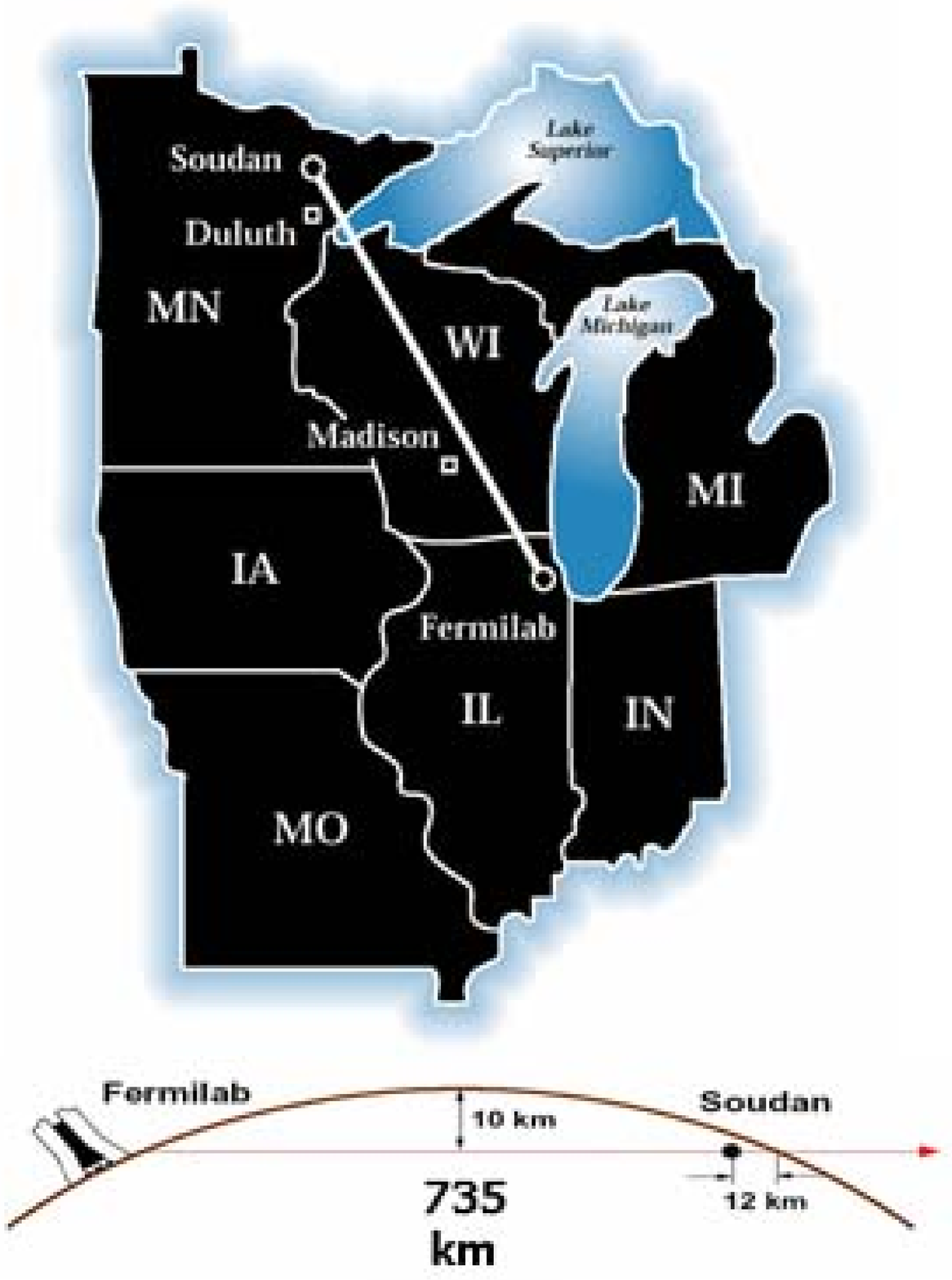}
\caption{\label{label} A scheme of the MINOS experiment.}
\end{minipage}\hspace{2pc}%
\begin{minipage}{18pc}
\includegraphics[width=18pc]{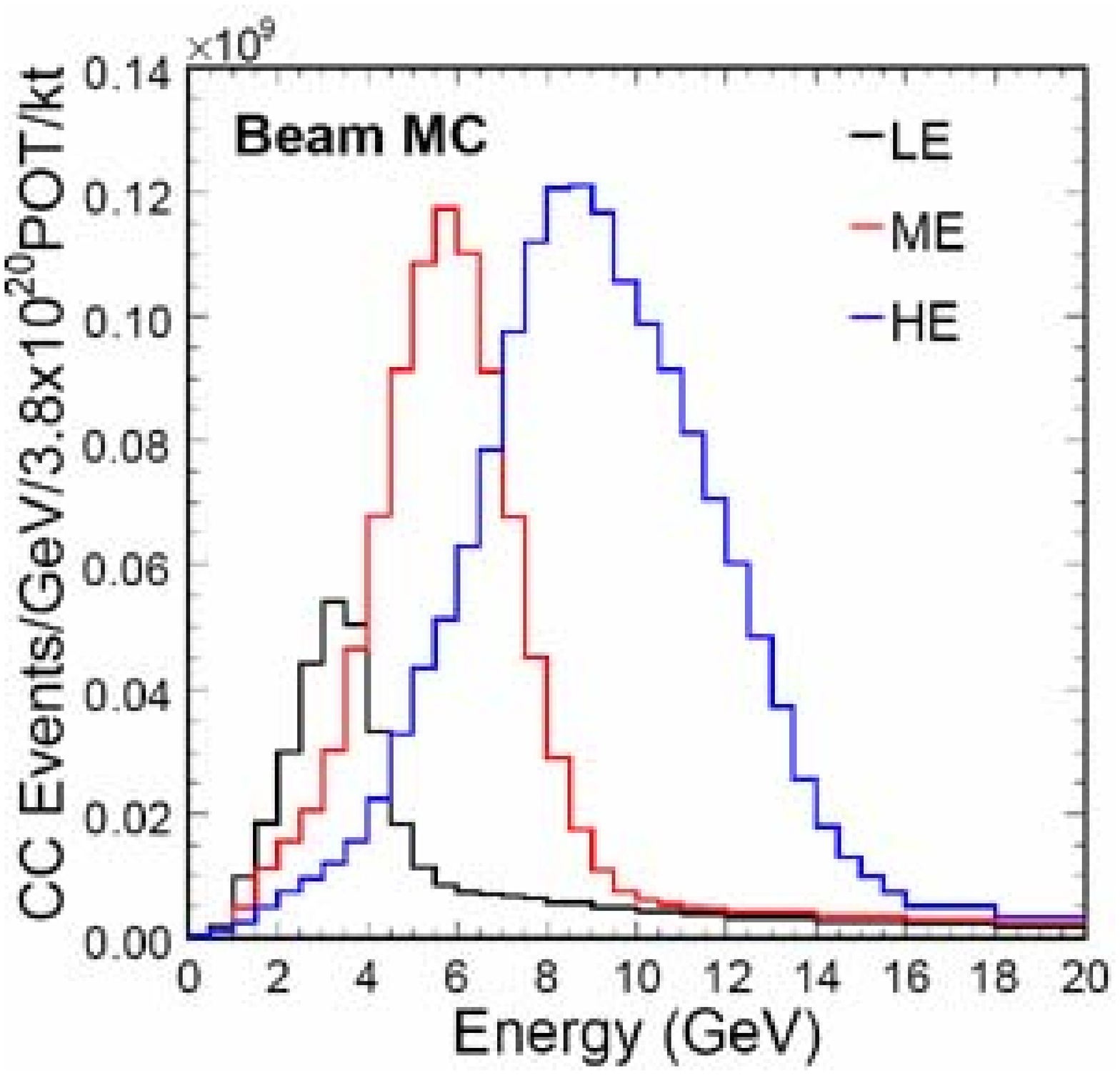}
\caption{\label{label} The three possible MINOS beams.}
\end{minipage}
\end{figure}

The experiment uses the NuMI (Neutrinos at the Main Injector) beam at the Fermi National Laboratory. The experiment has two detectors, a Near Detector located at FNAL and a Far Detector located at a distance of 735 km in the Soudan mine in Northern Minnesota (Fig. 11).
Both detectors have a similar design consisting in alternate planes of steel and plastic scintillator. The near detector has a mass of 1 kt and it is located at 1 km from the neutrino production point. The far detector has a mass of 5.4 kts and consists of two super-modules (SM) separated by a gap of 1.1m. The structure is that of a sampling calorimeter consisting of octagonal planes of 2.54 cm thick steel followed by planes of 1 cm thick scintillator and a 2 cm wide air gap. The first and second SMs contain 248 and 236 scintillator planes respectively. The SMs are magnetised to 1.5T by coils running  along the detector central axis and returning below the detector. The scintillator planes are made of strips 4 cm wide consisting on bars of scintillator that can be up to 8 m long depending on their position in the plane (the SciBar detector of K2K was build with the same type of bars).  The orientation of the strips alternate between $\pm 45^0$ with respect to the vertical from plane to plane. Each bar is read out by a scintillating fiber running through a hole in its center. 

\begin{figure}[h]
\begin{minipage}{18pc}
\includegraphics[width=18pc]{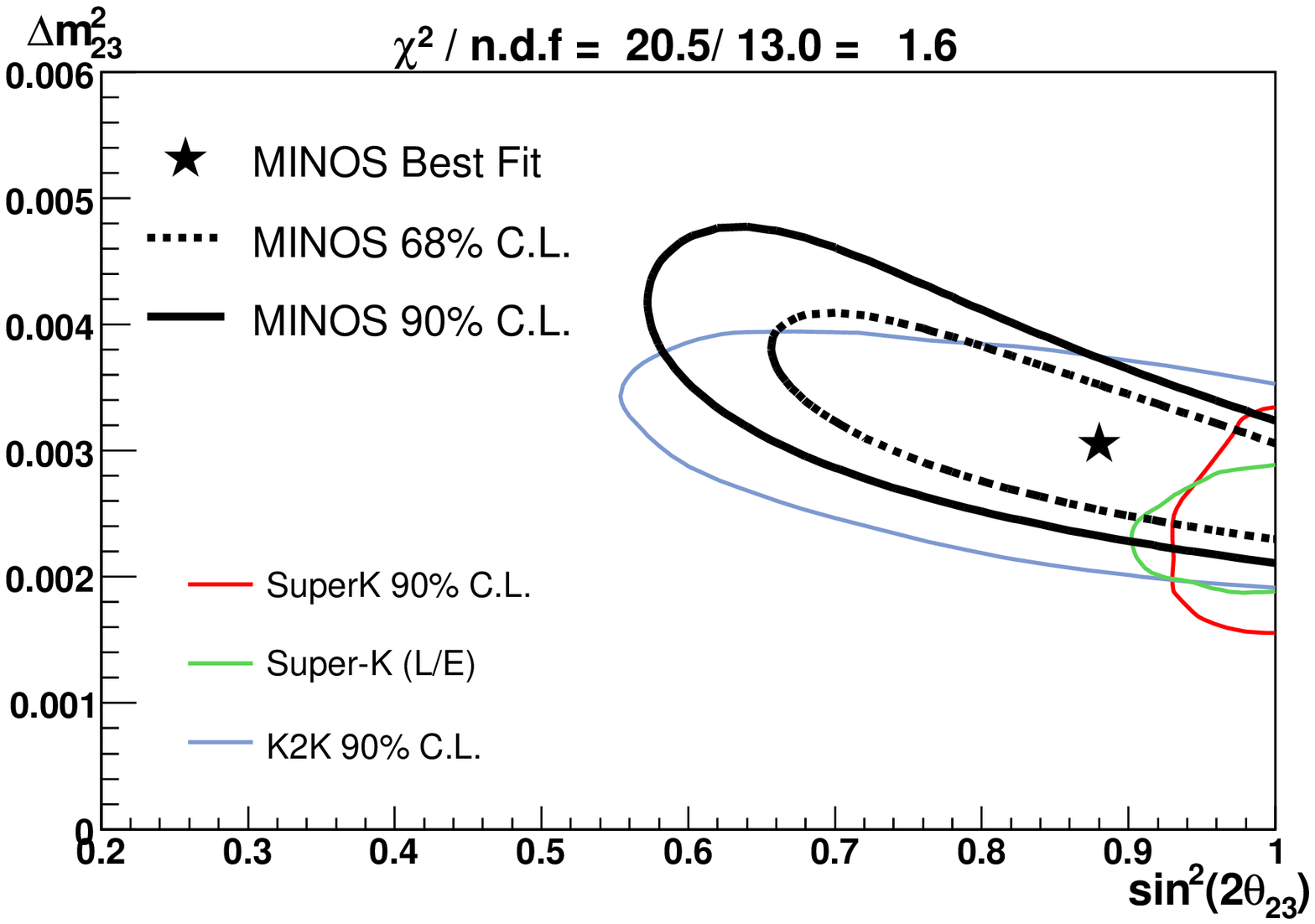}
\caption{\label{label} The allowed region of the $\Delta m^2_{23}$ versus $sin^2\theta_{23}$ plane obtained by MINOS \cite{minos-press}.}
\end{minipage}\hspace{2pc}%
\begin{minipage}{18pc}
\includegraphics[width=18pc]{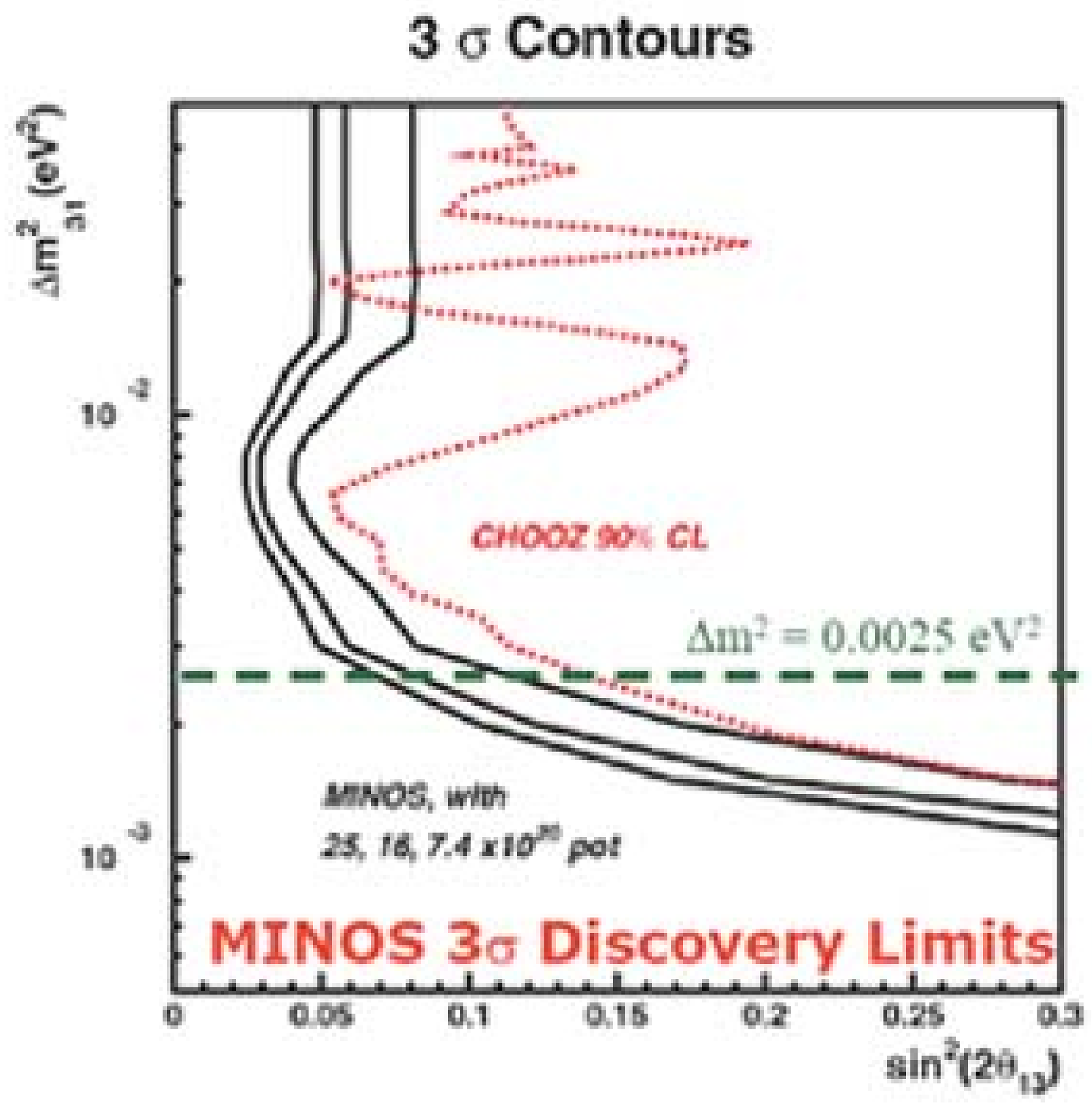}
\caption{\label{label} The expected MINOS $3\sigma$ discovery potential for a non-zero value of $\theta_{13}$ for three different numbers of protons on target \cite{thomson}.}
\end{minipage}
\end{figure}

The neutrino beam is produced with 120 GeV protons from the main injector, which are extracted every 1.9s with a spill of 8.7 $\mu s$. A system of two horns focuses positive pions produced in the collisions of the proton beam with a graphite target. The distance between the two horns and their position with respect to the target can be adjusted to produce three different beam profiles (Fig.12). For the $\Delta m^2_{\mu e}=2.5\times10^{-3} eV^2$ and a distance of 735 km the oscillation maximum corresponds to $E=1.3 GeV$ and therefore the low energy beam is that chosen for the initial goals of the experiment. It is expected that the accelerator will provide $2.5 \times 10^{20}$ protons on target per year, corresponding to a beam power of 0.3 MW.

One of the goals of MINOS will be to study the spectrum distortion due to $\nu_{\mu}$ disappearance. From this observation the goal is to measure $\Delta m^2_{23}$ to 10\% by the end of the experiment. Preliminary results \cite{minos-press} have already been presented (Figure 13). MINOS can also look for the subdominant $\nu_{\mu}$ to $\nu_e$ transition (see below). The final reach on the plane $\Delta m^2_{13}$ versus $\theta_{13}$ depends on the total number of protons on target and is illustrated in Figure 14. Another feature, since the experiment has a magnetized detector, is that it will be able to separate the neutrino and antineutrino components of atmospheric neutrinos and look for possible asymmetries due to matter effects \cite{minos1}.

\subsection{The CNGS Project}

The CERN to Gran Sasso (CNGS) project \cite{CNGS} includes another two long baseline experiments aiming at the confirmation of the atmospheric oscillations, this time by seeing explicitly the oscillation $\nu_{\mu}$ to $\nu_{\tau}$ through the identification of $\tau$ leptons produced by $\nu_{\tau}$ interactions in the far detector. These are therefore appearance experiments, unlike all the other experiments up to now. The baseline of the project is almost identical to that of MINOS, 732 km. Two experiments are foreseen: OPERA \cite{OPERA}, where the taus are identified in nuclear emulsions and ICARUS \cite{ICARUS}, based on a liquid Argon TPC. The beam is produced from protons of the CERN SPS and its energy is optimized for identification of $\nu_{\tau}$ interactions. The optimization involves two competing effects: the oscillation probability which is maximum at about 1.3 GeV for the CERN to GS distance, as we have already seen, and then decreases as $sin^2 (1.2\Delta m^2_{23}L/E)$ with increasing energy, and the probability of $\nu_{\tau}$ interactions, which grows linearly with $E_{\nu_{\tau}}$. It turns out that the optimal energy is at about 17 GeV, which is the design energy of the beam. The beam will be ready in 2006.

\begin{figure}[h]
\begin{minipage}{18pc}
\includegraphics[width=18pc]{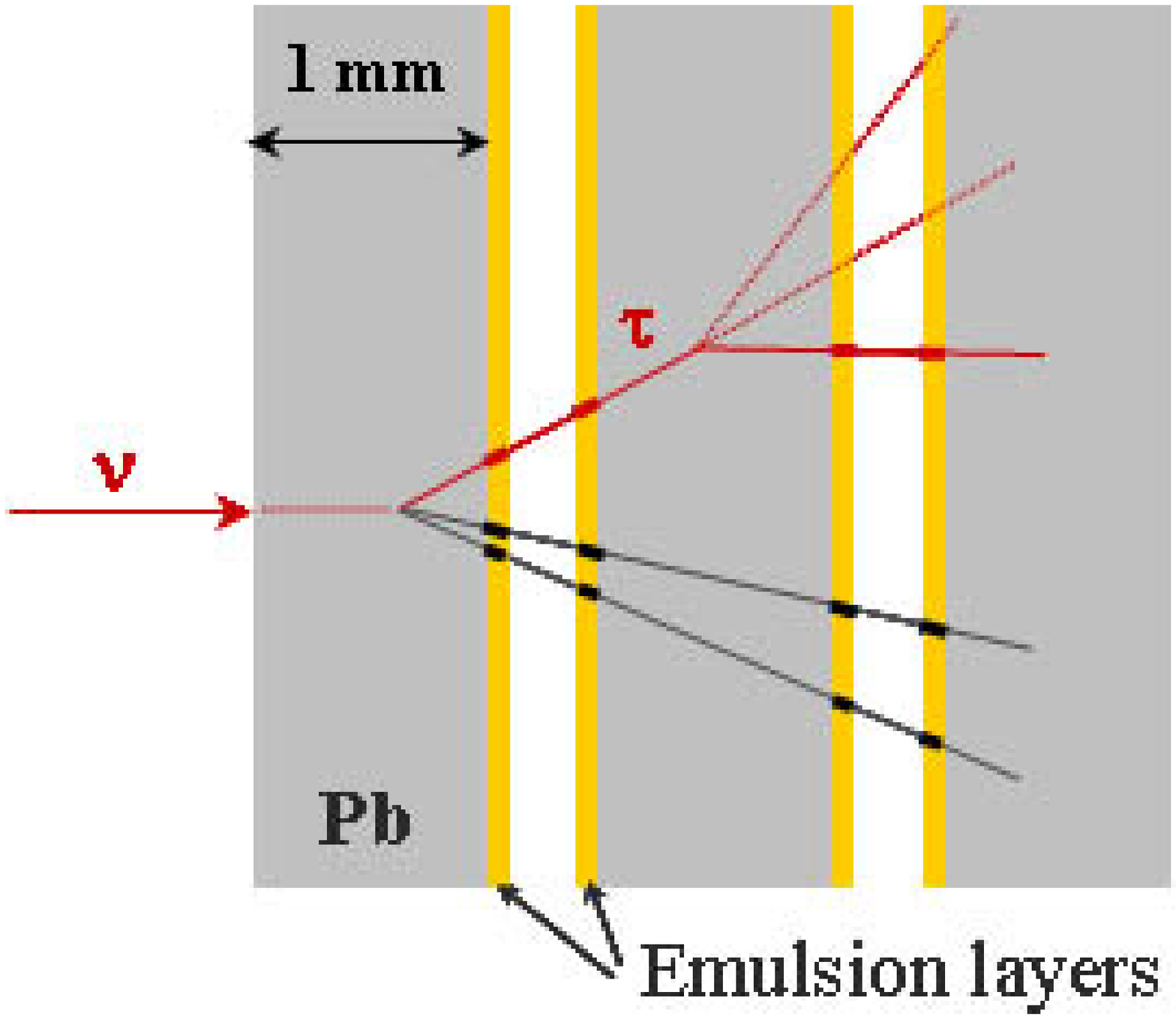}
\caption{\label{label} The detection principle of the OPERA experiment. The figure is from \cite{opera3}.}
\end{minipage}\hspace{2pc}%
\begin{minipage}{18pc}
\includegraphics[width=18pc]{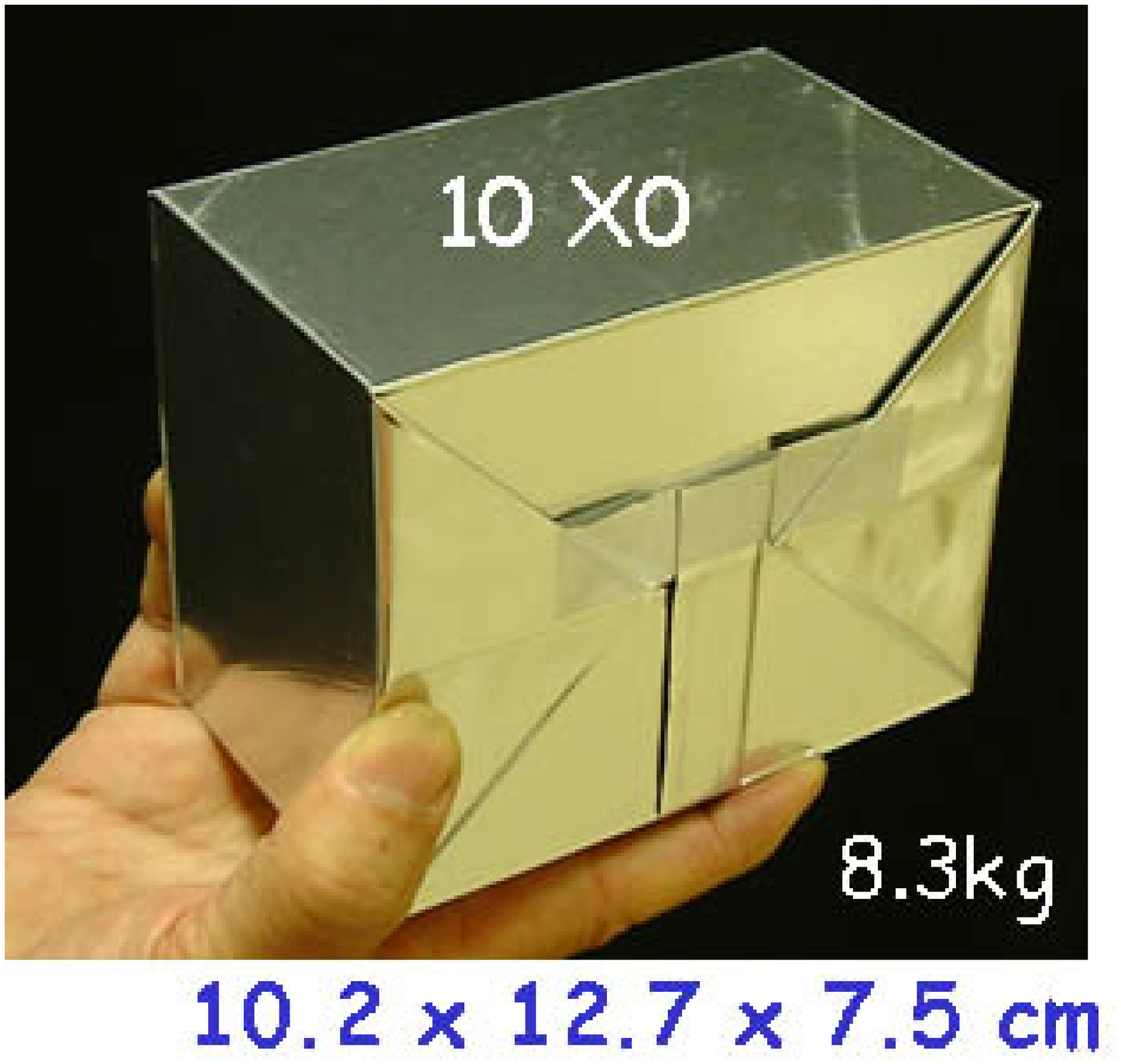}
\caption{\label{label} The basic unit of the OPERA detector. The figure is from \cite{opera3}.}
\end{minipage}
\end{figure}

The OPERA experiments searches for $\nu_{\tau}$ appearance by detecting the $\tau$ produced in $\nu_{\tau}$ charged current interactions. The aim is to confirm that the dominant oscillation in atmospheric neutrinos is $\nu_{\mu}$ to $\nu_{\tau}$ as implied by the Super-Kamiokande and K2K results. The $\tau$ detection is based in the identification of a track with a kink: a charged particle produced in the $\tau$ decay will not be pointing to the primary vertex of the interaction, reconstructed from the other tracks in the event (see Fig. 15). The method requires a detector with very good spatial resolution which in the case of OPERA is achieved with the use of emulsions. The basic detection unit is the so called ``brick" (Fig. 16). Each brick consists of a sandwich of 56 lead plates, 1mm thick, alternated with 56 emulsion sheets, weights 8.3 kg and has a thickness of about 10 radiation lengths. The primary vertex and the kink are detected with the information in the emulsions which also gives particle identification \cite{opera1}. The bricks are arranged in walls which alternate with hodoscopes of scintillator planes. The hodoscopes provide the trigger and localize the brick in which the interaction takes place, which is then extracted and scanned automatically. Following a set of 31 walls of bricks and scintillator planes there is a muon spectrometer. The set is known as a super-module and the whole OPERA detector consists of two such super-modules. After 5 years of running with a nominal beam of $4.5\times 10^{19}$ protons on target one expects between 10 and 20 $\tau$ identifications, depending on the value of $\Delta m^2_{23}$.

\begin{figure}[h]
\begin{minipage}{22pc}
\includegraphics[width=22pc]{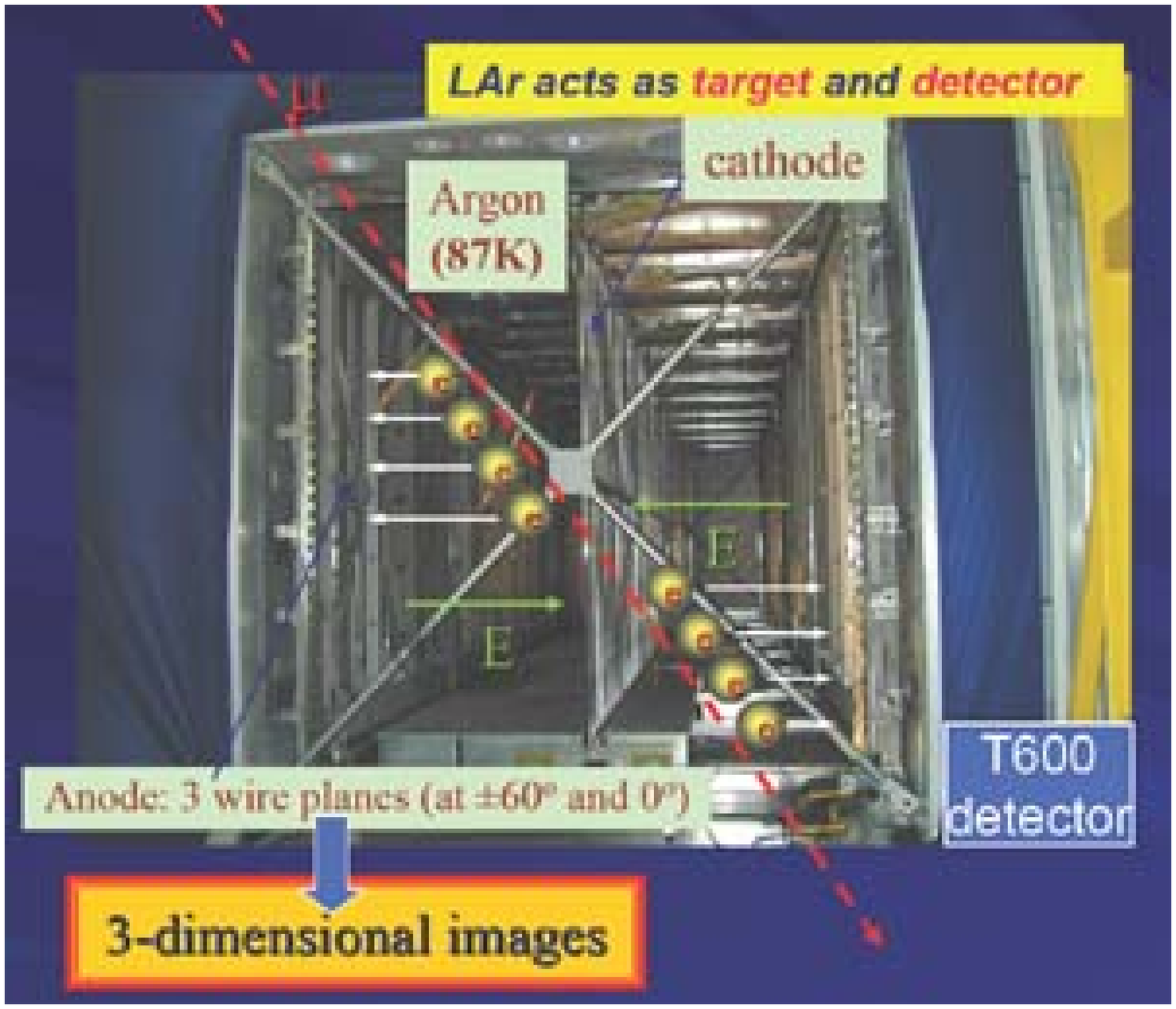}
\caption{\label{label} Illustration of the ICARUS principle of operation. The figure is from \cite{icarus3}.}
\end{minipage}\hspace{2pc}%
\end{figure}

ICARUS is based on the liquid Argon TPC idea, proposed already in the seventies \cite{rubbia}. The electrons produced in the ionization of charged tracks are drifted in the liquid towards the wall of the liquid argon tank (see Fig. 17) and detected in 3 anode wire planes oriented at 3 different angles to provide x-y coordinate in the plane which, together with the drift time, allows a 3-D reconstruction of the tracks. The collaboration has successfully operated the so-called T600 module which will be installed at the Gran Sasso \cite{icarus1}, \cite{icarus2}. The T600 operation has proven the technique, which provides very good spatial resolution and particle identification capabilities. However safety issues of storing large volumes of liquid Argon underground may limit the size of the detector with respect to the original proposal of installing up to 3000 tons of liquid Argon in the Gran Sasso. 

\section{Future oscillation experiments}
Despite the enormous progress in the understanding of the oscillations there are still some oscillation parameters for which we have no information. One is the value of the $\theta_{13}$ angle for which we only know that it is smaller than about $10^0$. The best hope for measuring this angle is to look for the disappearance of electron antineutrinos in short baseline reactor experiments or for the appearance of electron neutrinos due to the subdominant transition $\nu_{\mu}$ to $\nu_e$ in long baseline accelerator experiments. Two such experiments are proposed for the near future. One, already approved and under construction, is the T2K (Tokai to Kamioka) \cite{T2K} experiment in Japan, the other the NO$\nu$A experiment \cite{NOVA}, from FNAL to Ash River, Minnesota, in the US. 

We also do not know the sign of the $\Delta m^2_{32}$ mass difference, that is, we do not know if $m_3$ is larger or smaller than $m_1$ and $m_2$. This is known as the hierarchy problem. The first case, $m_3$ larger than the other masses, is know as normal hierarchy, while the latter case, $m_3$ smaller, is know as inverted hierarchy. Whether we have one case or the other is relevant for theories of neutrino masses. One hope for measuring the sign of the difference is to look for matter effects over sufficiently long distances in accelerator based experiments. Matter effects contribute with a term to oscillations which has a different sign for neutrinos and antineutrinos (and for normal and inverted hierarchy see equation 12 below). NO$\nu$A may have a chance to measure the mass hierarchy from this effect.

The other parameter completely unknown at present is the CP-violating phase $\delta$. This parameter can also be extracted from matter effects, but the ability to measure it in a given experiment will depend on the size of $\theta_{13}$. To measure this parameter will most likely require much more intense beams, as those mentioned in the next section.

In this section we concentrate in T2K and NO$\nu$A. A common characteristic of the beams for these experiments is the so called off-axis configuration: the direction of the beam is not aligned with that of the far detector, but at a small angle. The reason for this is that the energy of the neutrinos emerging at a small angle from the direction of the proton beam is roughly constant, independent of the primary proton energy (Fig. 18) \cite{off-axis}. This is a consequence of that fact that the pion, a scalar particle, decays isotropically in its rest frame and of the strong Lorenz boosts involved. The advantage of having these quasi-monochromatic beams is twofold. First the flux at the chosen energy (angle) is actually higher than that one would obtain with the standard on axis wide-band beam. This allows to optimize the beam at the energy of interest, e.g. at the oscillation maximum for the baseline of the experiment. Second, the suppression of the higher energies minimizes the background. For example in the case of T2K an important background comes from neutral current interactions producing a $\pi^0$, where one of the gammas from its decay is lost and the other is confused with an electron. The event appears then as a charged current electron neutrino interaction of lower energy. Minimizing this cross-section clearly helps for reducing this background. 

\begin{figure}[h]
\begin{minipage}{18pc}
\includegraphics[width=18pc]{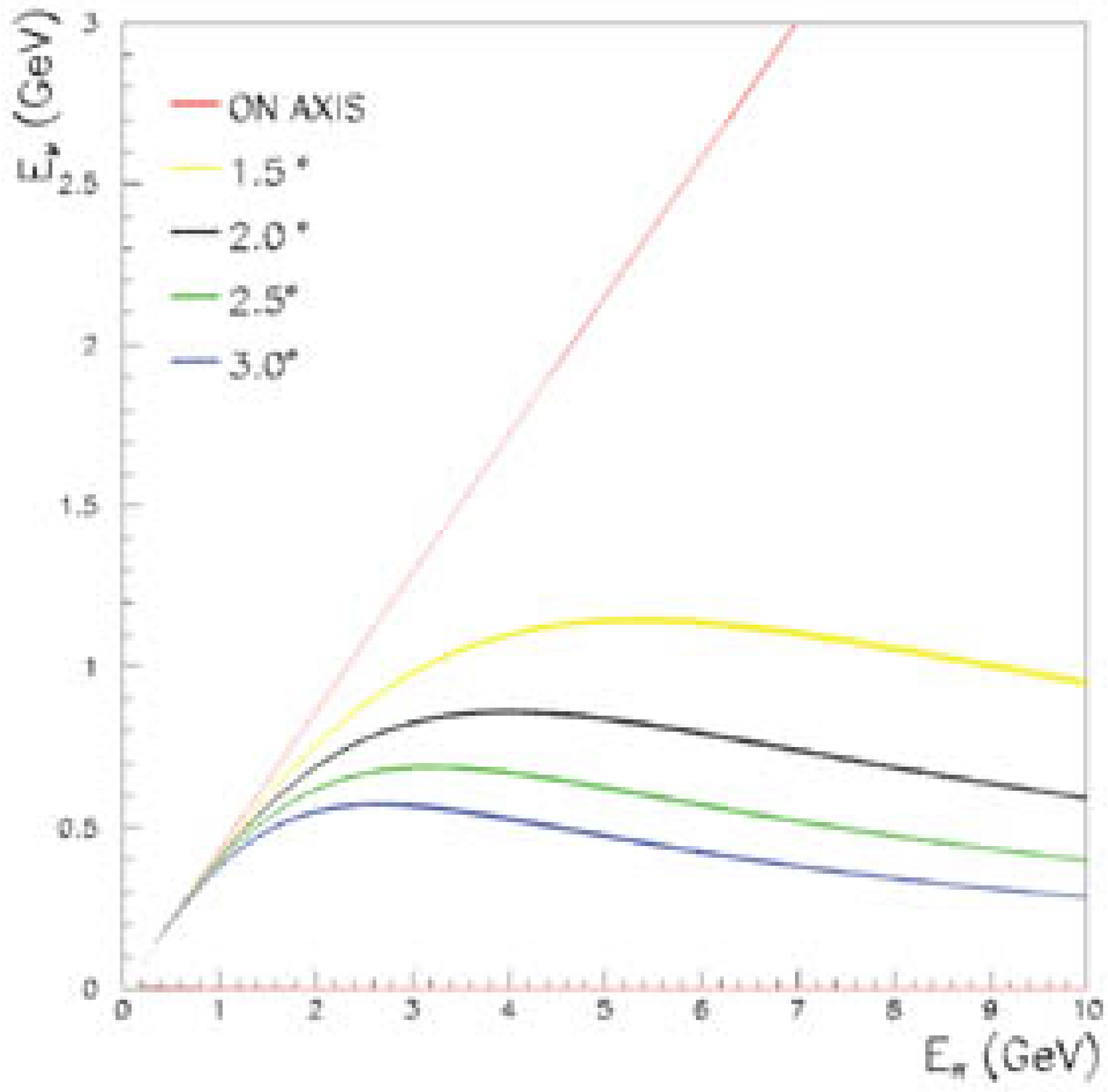}
\caption{\label{label} The energy of the neutrino as a function of the parent pion energy for several off-axis angles.}
\end{minipage}\hspace{2pc}%
\begin{minipage}{18pc}
\includegraphics[width=18pc]{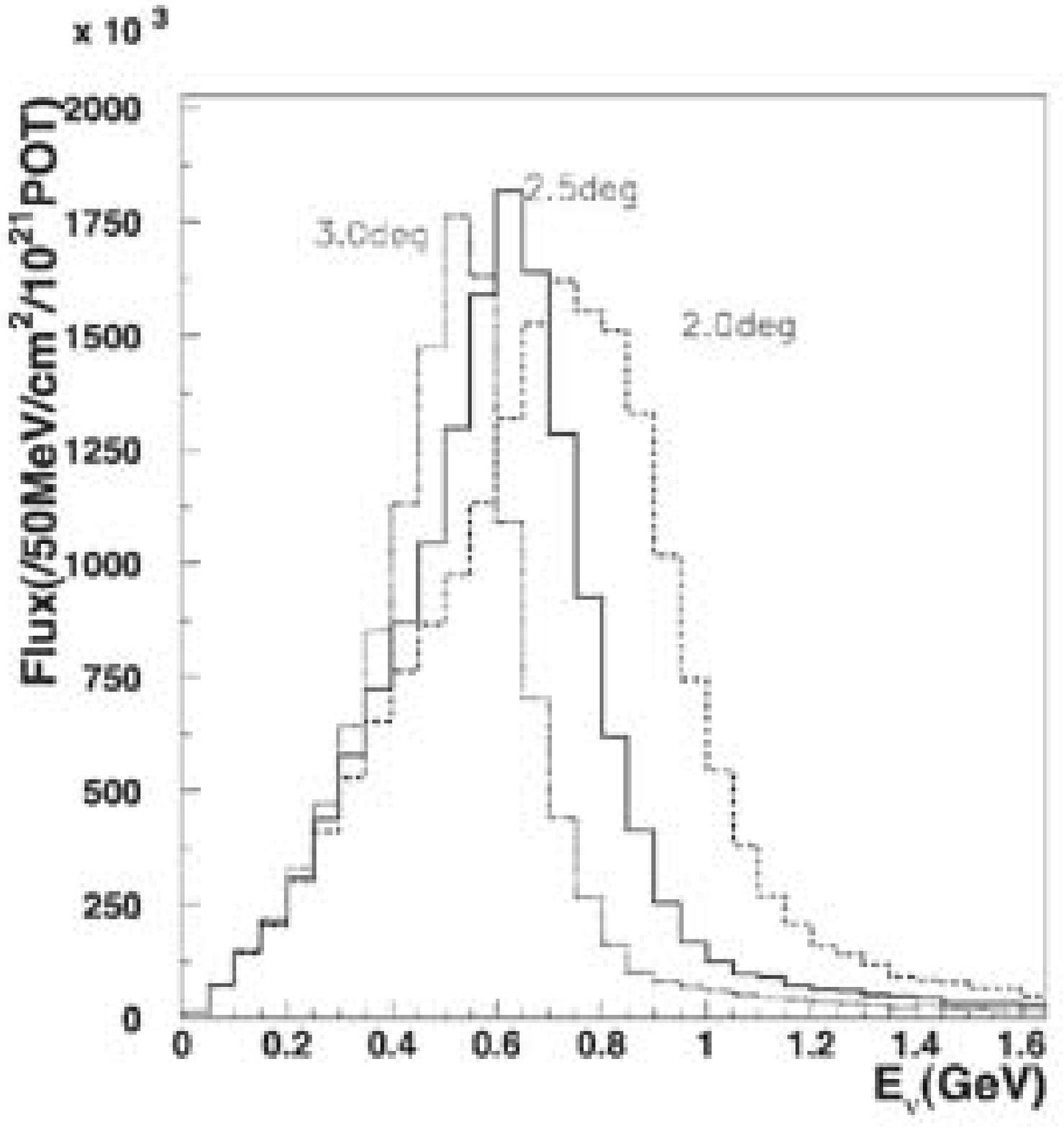}
\caption{\label{label} The energy distribution of the T2K beam for several off-axis angles.}
\end{minipage}
\end{figure}

As already mentioned T2K and NO$\nu$A will look for the appearance of electron neutrinos in a conventional muon neutrino beam. The probability for this transition is given by \cite{freund}
\begin{eqnarray} 
P(\nu_{\mu}\rightarrow\nu_e) & \approx & sin^2\theta_{13} sin^2\theta_{23} sin^2 \Delta \nonumber \\
& \mp & \alpha sin\delta cos\theta_{13} sin2\theta_{12} sin2\theta_{13} sin2\theta_{23} sin^3 \Delta \nonumber \\
& + & \alpha   cos\delta cos\theta_{13} sin2\theta_{12} sin2\theta_{13} sin2\theta_{23} cos\Delta sin^2 \Delta \nonumber \\
& + & \alpha^2 cos^2 \theta_{23} sin^2 2\theta_{12} sin^2 \Delta
\end{eqnarray}
\noindent Here $\Delta =\frac{\Delta m^2_{31} L}{4E_{\nu}}$ and $\alpha=\frac{\Delta m^2_{12}}{\Delta m^2_{31}}$. Since $\Delta m^2_{12}$ is smaller (in absolute value) than $\Delta m^2_{32}$, $\Delta m^2_{31}\approx\Delta m^2_{32}$. The - sign is for neutrinos the + sign for antineutrinos.

The above formula has some degeneracy, that is, different combinations of parameters can give rise to the same value of $ P(\nu_{\mu}\rightarrow\nu_e) $.
The overall degeneracy is eight-fold \cite{barger} and it is induced by the unknown sign of $\Delta m^2\theta_{31}$, by the correlation between $\theta_{13}$ and $\delta$ and by the fact that the parameter $\theta_{23}$ is known with the ambiguity between $\theta_{23}$ and $\pi/2 - \theta_{23}$ (this ambiguity disappears for $\theta_{23}=\pi/4$, the preferred value from the fits).

The disappearance probability for electron antineutrinos, relevant for reactor experiments, is given by
\begin{eqnarray}
1-P(\bar{\nu}_e \rightarrow \bar{\nu}_e) \approx sin^2 2\theta_{13} sin^2 \Delta + \alpha^2 \Delta^2 cos^4 \theta_{13} sin^2 2\theta_{12}
\end{eqnarray}
\noindent From this last formula it is clear that reactor experiments are not sensitive to the CP violating phase nor to the sign of the mass difference, but for this reason they can provide a measurement of $\theta_{13}$ free from ambiguities. Several experiments are now proposed with the aim of measuring $\theta_{13}$ or improve the Chooz upper limit \cite{oberauer}.

\subsection{T2K}
T2K (Tokai to Kamioka) \cite{T2K} is the name of an experiment which will use a neutrino beam produced by a 50 GeV proton synchrotron in construction in Tokai, named J-Park, directed towards the Super-Kamiokande detector. The configuration will be off-axis, with a beam that can be steered such that the Super-Kamiokande detector can be at between 2 and 3 degrees off axis, corresponding to energies between 0.5 and 0.9 GeV (Fig. 19). For $\Delta m^2_{23}=2.5\times10^{-3} eV/c^2$, an $E_{\nu}=0.6 GeV$ the first oscillation maximum is at about 295 km, the distance from Tokai to Kamioka.

\begin{figure}[h]
\begin{minipage}{18pc}
\includegraphics[width=18pc]{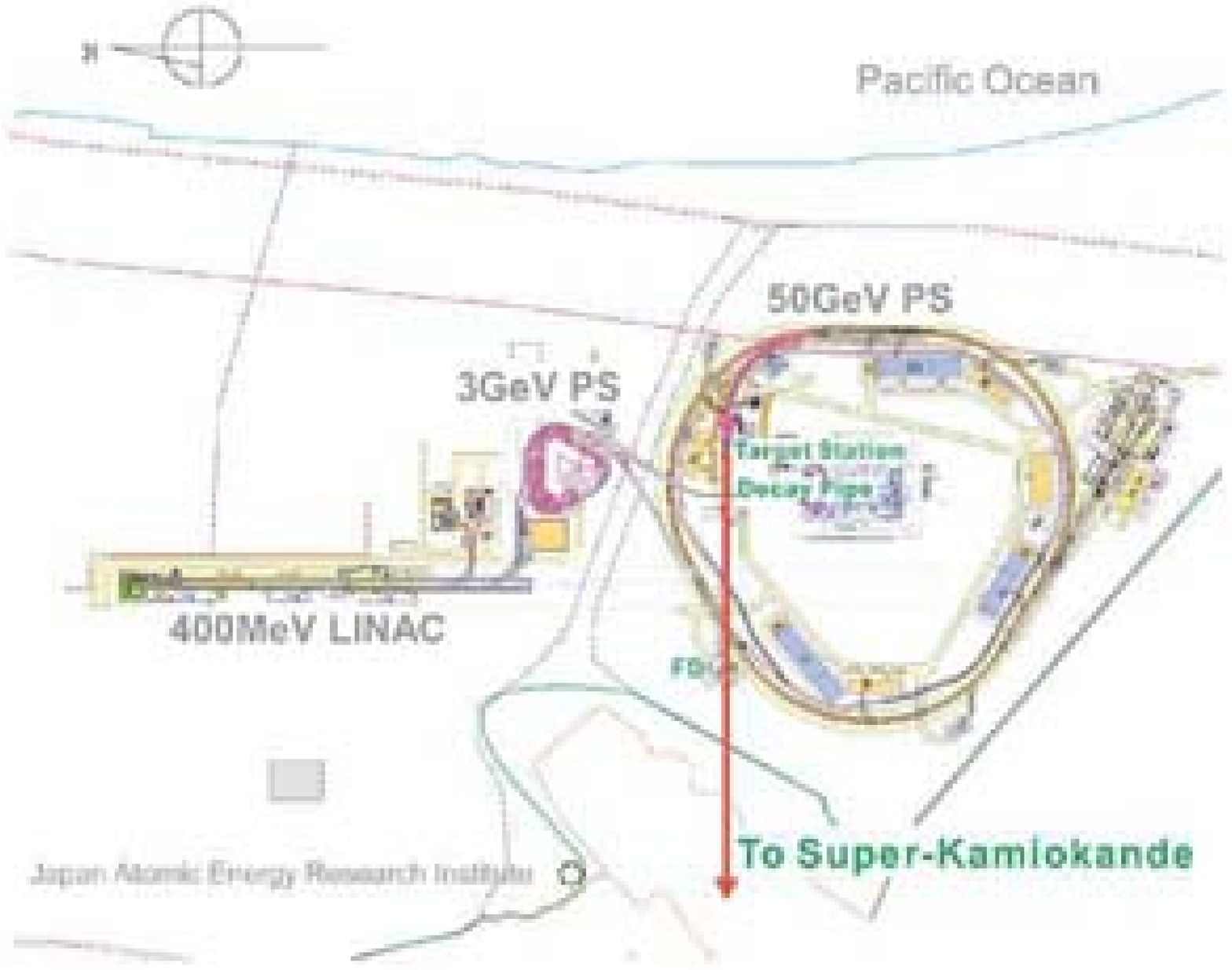}
\caption{\label{label} The J-PARK accelerator complex and the T2K neutrino beam.}
\end{minipage}\hspace{2pc}%
\begin{minipage}{18pc}
\includegraphics[width=18pc]{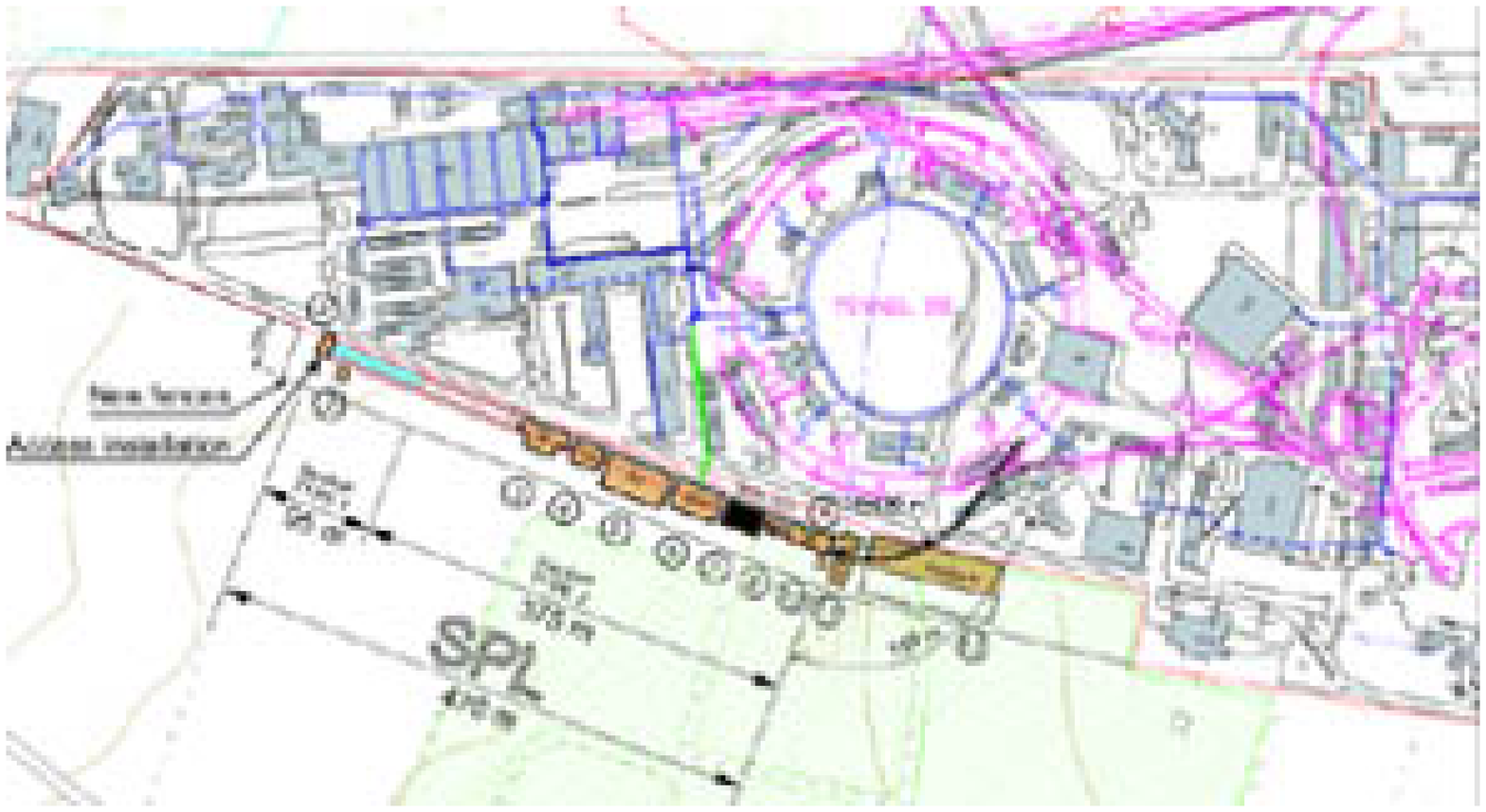}
\caption{\label{label} A scheme of the SPL layout at CERN.}
\end{minipage}
\end{figure}

In the first phase of the experiment, schedule to start in 2009, the beam power will reach 0.75 MW. During this phase the experiment will focus on searching for the subdominant oscillation $\nu_{\mu}\rightarrow\nu_e$. T2K will have a near-detector complex at 280 m from the pion production target, which is located on the inner side of the J-PARK ring (Fig. 20). There are two detectors at 280 m: an on-axis detector consisting of 13 modules placed in a cross-configuration with the purpose of measuring precisely the neutrino beam direction, and an off-axis detector to measure the energy spectrum of muon neutrinos and to estimate the contamination of electron neutrinos, which is estimated as 0.4\%. This detector will be sitting inside a large magnet (used previously in the UA1 and NOMAD experiments at CERN) and therefore it would be possible to separate the muon neutrino and antineutrino components in the beam. 

The neutrino beam energy will be at the maximum of the oscillation probability for $\nu_{\mu}$ to $\nu_{\tau}$ for which the dominant reaction is the quasi-elastic interaction 
\begin{eqnarray} \nu_l + N \rightarrow l + N^{'} \end{eqnarray}
\noindent for which the neutrino energy can be reconstructed from the energy and angle of the outgoing lepton, as we have seen in section 4.1, equation (11). The energy resolution is of the order of 10\%.
Muon and electron neutrino events can be easily distinguished in the far detector, Super-Kamiokande. The main problem comes from the neutral current $\pi^0$ production where one of the gammas from the $\pi^0$ decay is lost and the other can be confused with an electron. Since the experiment runs at low energy the $\pi^0$ production cross-section is greatly reduced. When proper cuts are applied the rejection efficiency is expected to reach 99\%, with an efficiency for retaining the signal events of the order of 40\%. For the purpose of studying the physics reach of the experiment it is assumed that one would get $10^{21}$ protons on target per year, with a primary proton beam energy of $40 GeV$. The estimated sensitivity for $sin^22\theta_{13}$ is 0.008 at 90\% CL.

A second objective of the experiment is a precise determination of $\Delta m^2_{32}$ and of $\theta_{23}$ from measuring the probability of $\nu_{\mu}$ disappearance. The expected errors on the determination of these parameters will be about $10^{-4}$ and 0.01 respectively.

\subsection{NO$\nu$A}

NO$\nu$A will use the NUMI beam line with a far detector off-axis located at 810 km from FNAL in Ash River, Minnesota. The experiment also aims at measuring $\theta_{13}$ similarly to T2K with similar value of L/E (and therefore higher neutrino energy). At such energy and distance matter effects are non negligible and therefore the experiment has some sensitivity to the sign of $\Delta^2m_{32}$ and the CP violating phase $\delta$.

NO$\nu$A will also have two detectors with similar structure. The near detector will measure the beam composition and its energy spectrum. The planned far detector is a 30 kt ``all active" tracking calorimeter made from planes of extrusion filled with liquid scintillator. The sensitivity to $\theta_{13}$ depends on the value of $\delta$, on the mass hierarchy and on the value of $\Delta m^2_{23}$. Assuming 5 years of running at $6.5 \times 10^{20}$ protons on target/year the sensitivity is better than 0.01 in most scenarios. The reader is referred to the NO$\nu$A proposal \cite{NOVA} for further information, including the discussion on how to extract the mass hierarchy and possibly $\delta$. The synergy of NO$\nu$A with T2K and other possible measurements is also discussed in \cite{parke}.

\begin{figure}[h]
\begin{minipage}{18pc}
\includegraphics[width=18pc]{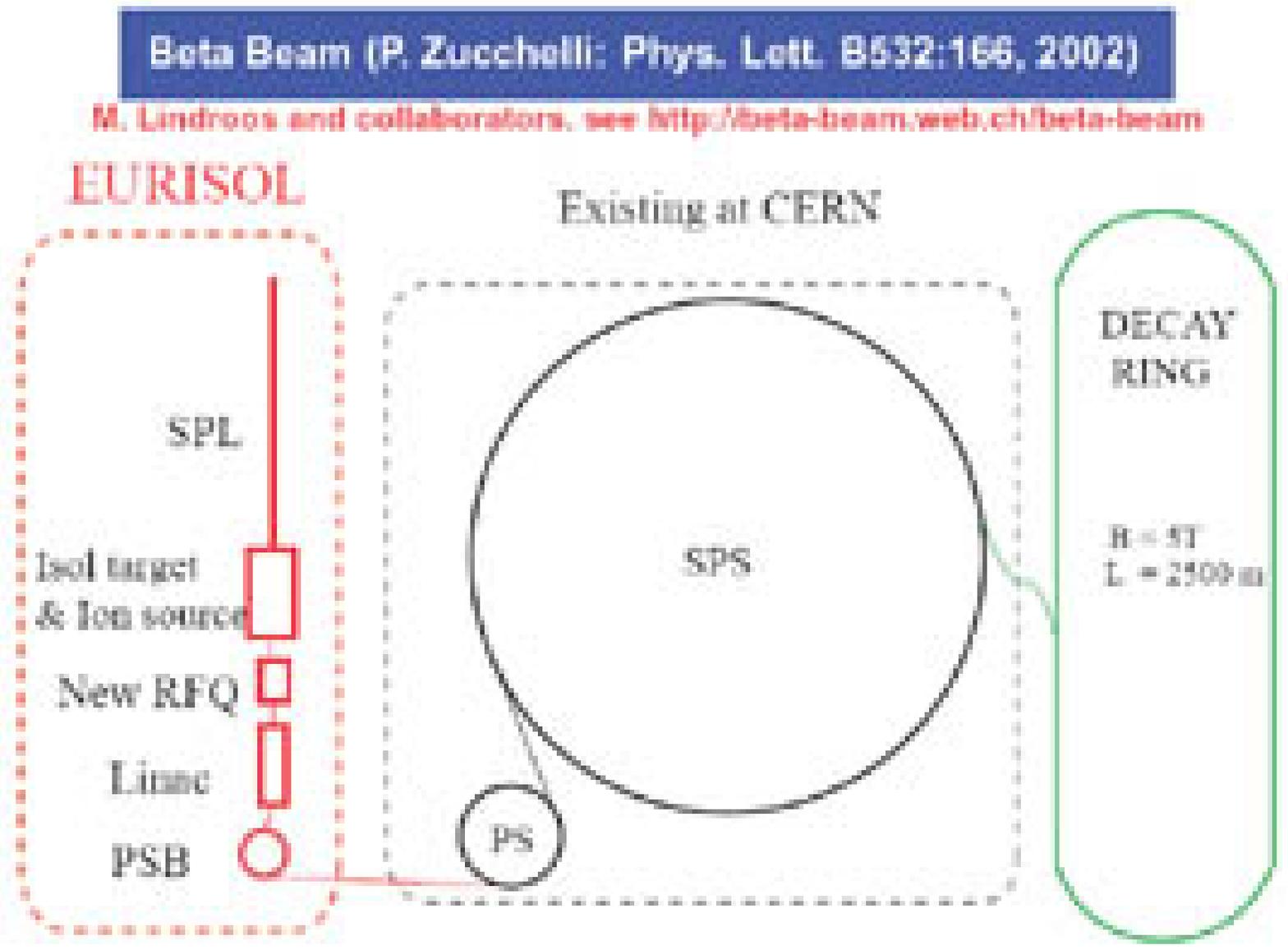}
\caption{\label{label} A scheme of the beta-beam facility at CERN.}
\end{minipage}\hspace{2pc}%
\begin{minipage}{18pc}
\includegraphics[width=18pc]{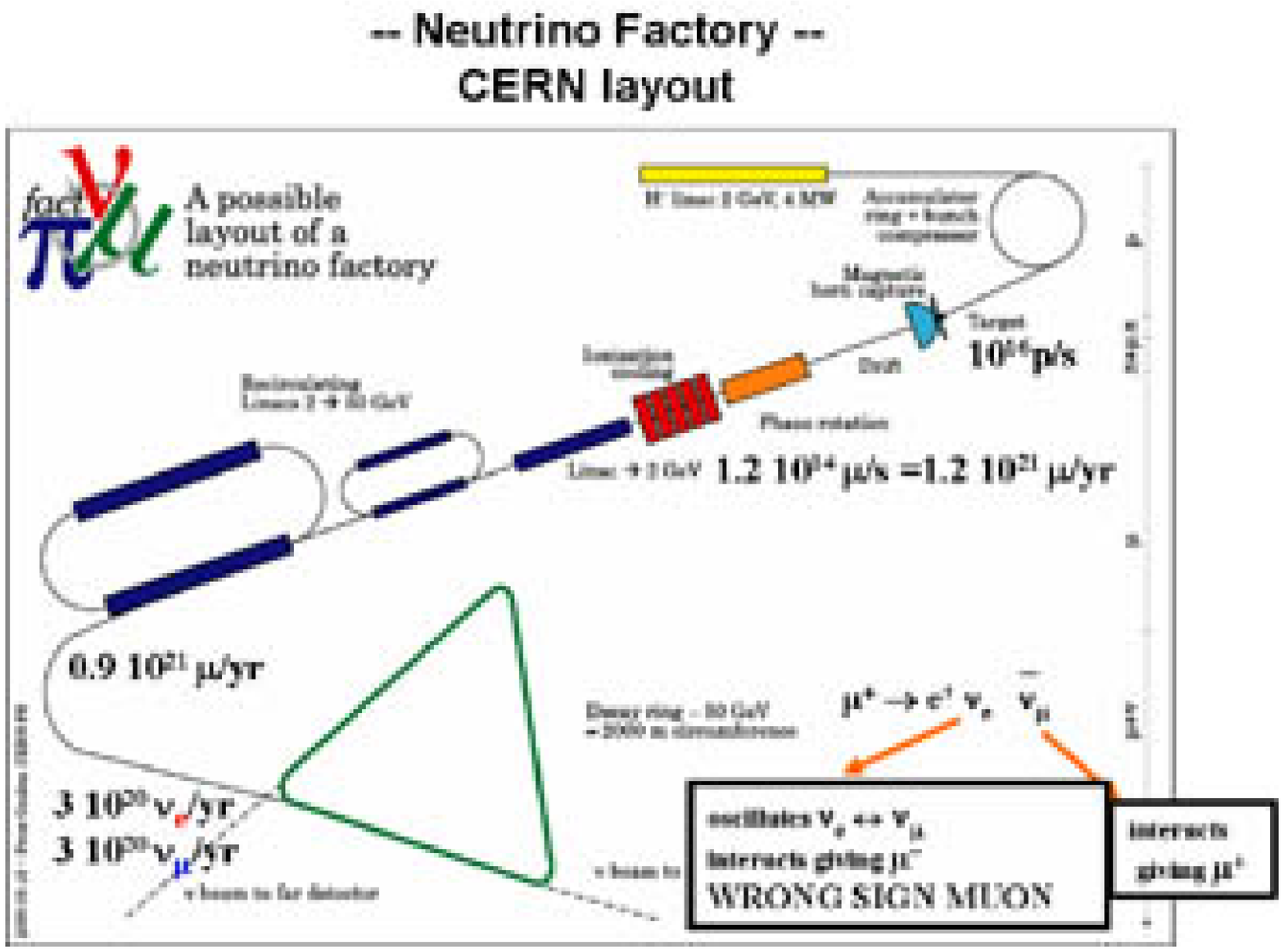}
\caption{\label{label} A scheme of the Neutrino Factory.}
\end{minipage}
\end{figure}

\subsection{Future Super-Beams}

In the context of neutrino experiments super beams are conventional beams (that is, with neutrinos produced in the decay of horn-focused $\pi$ and $K$ mesons, produced in proton-target interactions) but more intense than what has been possible up to now, involving primary proton intensities corresponding to about 1MW power or above. It is foreseen that the T2K beam be upgraded eventually to 4 MW power. For this phase a very large detector, Hyper-Kamiokande, a one Megaton water Cherenkov detector, sitting in a different location than that of the present Super-Kamiokande, is being proposed. The NUMI beam could also be upgraded to super-beam class, by building a new Proton Driver at FNAL. 

Another project for a superbeam is the SPL project at CERN \cite{SPL}, \cite{spl-web}. The SPL (Super Proton Linac) is a superconducting $H^{-}$ linear accelerator at CERN (originally intended as a proton accelerator and hence the name) which will produce a 2.2 GeV beam with a power of 4 MW. The facility will use most of the superconducting RF cavities available after the decommissioning of LEP, and will deliver $10^{16}$ protons per second, in 2.2 ms bursts with a repetition rate of 75 Hz. It can be located inside the CERN-Meyrin site (Fig. 21). The facility will have many purposes: at an early stage, it will upgrade the performance of the PS complex by replacing Linac2 and the PS Booster, by injecting protons directly into the PS. This will allow the construction of a conventional neutrino superbeam based on pion and kaon decay. If built, the SPL would have synergies with several CERN programs: the brilliance of the LHC accelerator could be tripled, the ISOLDE facility could be supplied with five times more beam than at present and the fixed target program based on the PS-SPS complex would also benefit. The SPL could also be used in the first stage of a new-generation radioactive ion beam facility (see below) and in conjunction with an accumulator and a compressor it would serve as the proton driver of a neutrino factory (see below). 

The physics of the SPL has been discussed in connection with a very large detector at the Modane Laboratory located in the Fr\'ejus tunnel, at 130 km distance \cite{mezzetto}, \cite{memphys}. The detector, called MEMPHYS, will be a 440 kt Water Cherenkov. The plan is to use this detector with a superbeam from the SPL plus a Beta-beam (see below) also from CERN.

\subsection{Beta-Beams and Neutrino Factories}

Conventional neutrino beams, for example of $\nu_{\mu}$ neutrinos produced focusing positive pions, have intrinsic backgrounds of $\bar{\nu}_{\mu}$, $\nu_e$ and $\bar{\nu}_e$, and moreover both the flux and these intrinsic backgrounds are difficult to predict with precision. Possible ways to overcome these limitations are Beta-Beams and Neutrino Factories. Here a very short account is given of these new ideas and the reader is referred to the already abundant literature.

\begin{figure}[h]
\begin{minipage}{25pc}
\includegraphics[width=25pc]{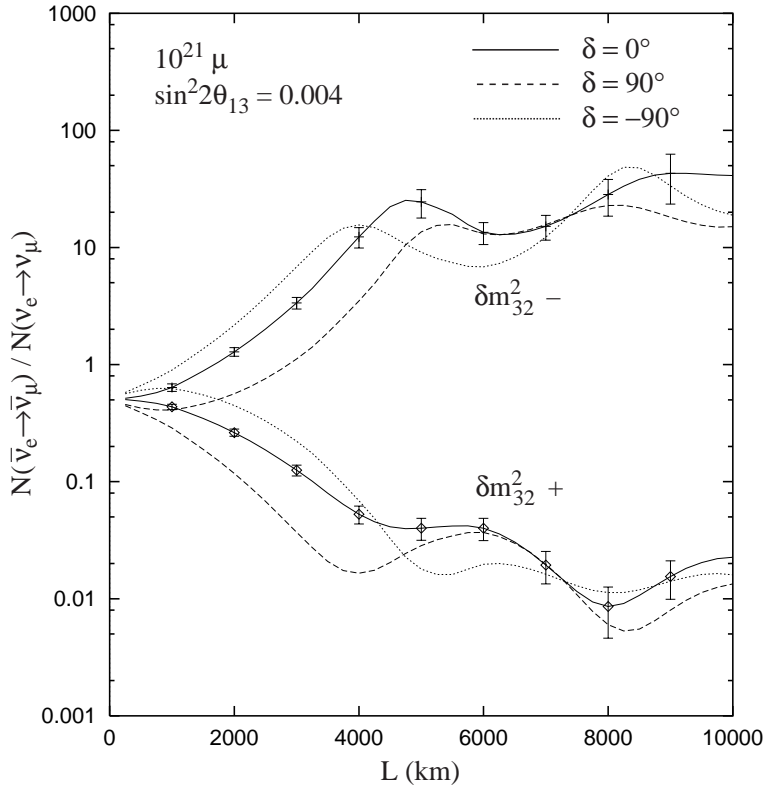}
\caption{\label{label} The ratio of the oscillation probabilities of neutrinos and antineutrinos as a function of the baseline. See text for an explanation \cite{apollonio}, \cite{barger2}.}
\end{minipage}\hspace{2pc}%
\end{figure}
 
Beta-Beams were introduced by P. Zucchelli in 2002 \cite{zucchelli}. The idea is to produce intense beam of $\nu_e$ or $\bar{\nu}_e$ from the decay of well collimated radioactive ions, beta+ or beta- emitters. The ions will be produced, collected, accelerated and stored in a decay ring. The ideal ions would have a lifetime not too short or two long, of the order of 1 s, and would be easily produced, such as $^6He$ and $^{18}Ne$. One specific study at CERN can be found in \cite{beta-cern} (Fig. 22). The project has a synergy with the Eurisol facility under study for CERN. The physics of beta-beams is discussed in \cite{mezzetto2}, \cite{memphys}, \cite{juanjo}, and references therein.

The Neutrino Factory idea was born from that of the Muon Collider \cite{mu-collider}, an accelerator complex capable of collecting, cooling, accelerating and storing counter-circulating beams of $\mu^+$ and $\mu^-$ at very high energies. That collider will produced very intense beams of neutrinos, that intense as to pose a radiation risk. It was soon realized that one could make intense neutrino beams from the decaying muons \cite{geer} which could be used to do interesting neutrino physics \cite{geer}, \cite{gavela}. For that purpose positive or negative muons will be collected, accelerated and injected in a storage ring with long straight sections, where they will let to decay, producing a forward intense beam of muon antineutrinos or neutrinos respectively. 

Studies for a Neutrino Factory complex have been carried out in Europe \cite{nufact-eu}, the US \cite{nufact-us} and Japan \cite{nufact-jp}. An International Scoping Study has been initiated in 2006 \cite{nufact-iss}. The reader is referred to \cite{apollonio}, \cite{burguet}, \cite{blondel} and references therein for the physics of the Neutrino Factory. Fig. 23 is an illustration of a possible Neutrino Factory layout. 

With these new kind of beams one could attempt to measure $\theta_{13}$, the mass hierarchy and, if $\theta_{13}$ is not too small, CP violation in the leptonic sector. In general it is necessary to run with neutrinos and antineutrinos to resolve all these problems free from ambiguities. In particular CP violation can be directly measured from the asymmetry
\begin{eqnarray}
\frac{P(\nu_{\mu}\rightarrow\nu_e)-P(\bar{\nu}_{\mu} \rightarrow \bar{\nu}_e)} 
{P(\nu_{\mu}\rightarrow\nu_e)+P(\bar{\nu}_{\mu} \rightarrow \bar{\nu}_e)}
\end{eqnarray}
\noindent The difference between the oscillation probabilities of neutrinos and antineutrinos as a function of the baseline, the mass hierarchy and the phase $\delta$, for a given value of $sin^22\theta_{13}$, is illustrated in Figure 24, taken from \cite{apollonio}, which was reproduced from \cite{barger2} for an energy of the muon beam of 30 GeV, $10^{21}$ muons decays for each charge and a 40 kt detector. It can be seen that the difference is maximal at distances between 2500 km and 3500 km. For this kind of measurement one would like to have a very long baseline, such as for example from CERN to a detector in the Canary Islands, at 2800 km distance. 

\section{Acknowledgements}
It is a pleasure to thank the organizers for the invitation to talk at such a good school in such a beautiful place, particularly George Fanourakis, Alberto Ruiz, Aurore Savoy-Navarro and George Zoupanos. Work is supported by the Spanish MEC, FPA2003-06921 project.

\end{document}